%% file: twist3.tex
\documentclass[a4paper,11pt]{article}
\usepackage[utf8]{inputenc}
\usepackage{graphicx}  
\usepackage{dcolumn}   
\usepackage{bm}        
\usepackage{amssymb}   
\usepackage{amsmath}
\usepackage{a4wide}
\usepackage{subfig}
\usepackage{amsfonts}
\usepackage{slashed}
\usepackage{appendix}
\usepackage{color}
\usepackage{hyperref}
\usepackage{cleveref}
\usepackage{tikz}
\usepackage{placeins}
\usepackage[normalem]{ulem}
\title{Charmonium content of $\chi_{c1}(3872)$ in light-cone sum rules at twist 3}
\author {T.~Akan$^1$\thanks{tarik.akan@bozok.edu.tr}, M.A.~Olpak$^2$\thanks{maolpak@thk.edu.tr}, A.~Özpineci$^3$\thanks{ozpineci@metu.edu.tr}}
\date{
$^1$Department of Physics, Yozgat Bozok University, Yozgat, Türkiye\\
$^2$Department of Electrical and Electronics Engineering, University of Turkish Aeronautical Assocation, Ankara, Türkiye\\
$^3$Department of Physics, Middle East Technical University, Ankara, Türkiye
}
\newcommand\rsout{\bgroup\markoverwith
  {\textcolor{red}{\rule[.5ex]{1pt}{1pt}}}\ULon}

\begin{document}

\maketitle

\begin{abstract}
 In this study, we utilize light-cone QCD sum rules at twist-3 accuracy to compute the coupling parameters of the $\chi_{c1}
 (2P)$ state with $D$ and $D^*$ mesons. 
 The analysis reveals that the observed $\chi_{c1}(3872)$ 
 meson incorporates significant 
 amounts of both charmonium and molecular components. 
 The interplay between these components highlights the exotic nature of $\chi_{c1}(3872)$.
 Furthermore, implications for a possible mixing between the $\chi_{b1}(2P)$ state and a potential $B\bar{B}^*$ molecule is discussed.
  These findings contribute to the understanding of exotic hadron states and their intricate internal structures.
\end{abstract}

\input{introduction}
\input{sum_rules_part}

\input{q2_correction}
\input{conclusion}
\newpage
\input{appendix}

\FloatBarrier
\bibliographystyle{ieeetr}
\bibliography{twist3.bib}

\end{document}

%% file: introduction.tex
\section{Introduction}
In the past, hadrons have been examined within the context of the (naive) quark model, which states that mesons are quark-antiquark bound states and baryons are color neutral bound states of three quarks.
Currently, it is an established fact that not all hadrons fit into this picture.
 For example, some meson states are not simply two-particle (quark-antiquark) states.
 They might consist of more intricate structures, or so-called exotic states, such as meson molecules, tetraquarks, etc.
 A well-known instance is $\chi_{c1}(3872)$ (or known as $X(3872)$), which has quantum numbers $J^{PC} = 1^{++}$.
 There are models that categorize $\chi_{c1}(3872)$ as a $DD^*$ molecule \cite{Kalashnikova_2019}, whereas other models discuss both molecular and charmonium contributions to the content of $\chi_{c1}(3872)$ \cite{Chen:2022asf,Gens:2021wyf,ramos2020molecular,zhao2021study}.
For example, consider the following ratio, which represents the partial width of the $\chi_{c1}(3872) \to \psi(2S)\gamma$ decay — first detected via the $ B^+ \to \chi_{c1}(3872) K^+ $ process — to that of the $\chi_{c1}(3872) \to J/\psi \gamma$ decay: 
\begin{equation}  
    R_{\psi\gamma}\equiv \dfrac{\Gamma_{\chi_{c1}(3872)\rightarrow \psi(2S)\gamma}}{\Gamma_{\chi_{c1}(3872)\rightarrow J/\psi \gamma}}.  
\end{equation}  
The large measured value for the $R_{\psi\gamma}$ ratio \cite{Aaij2023} raises doubts about the pure $DD^*$ molecular hypothesis for $\chi_{c1}(3872)$
\cite{Chen:2022asf,Gens:2021wyf,ramos2020molecular,zhao2021study}, unless certain assumptions are made \cite{nieves2012heavy}.
 Instead, it supports predictions from several other hypotheses, including conventional charmonium \cite{barnes2004charmonium,Zhang:2024fxy}, $c\bar c q\bar q$ tetraquark \cite{Hanhart:2025bun}, and mixed molecules with a significant compact component \cite{Cincioglu2020,Takeuchi2021,brambilla2024naturechic1left3872righttccleft3875right,Berwein_2024}.

In \cite{Cincioglu2016}, a model is developed which describes $\chi_{c1}(3872)$ as a superposition 
of a $\chi_{c1}(2P)$ component and a $DD^*$ molecule. One of the important parameters in the model 
is the coupling of $\chi_{c1}(2P)$ to $DD^*$ component. This coupling should either be fit to
experimental data, or calculated theoretically. At the moment, the available experimental data is not
precise enough to determine its precise value \cite{Cincioglu2016}.
QCD sum rules \cite{SHIFMAN1979385} or its enhanced version, light cone QCD sum rules \cite{BALITSKY1989509}, is one of the nonperturbative techniques 
that can be used for examining the coupling parameters in question \cite{Colangelo2000}.
In the study of this coupling constant, light cone distribution amplitudes of $\chi_{c1}(2P)$ are required \cite{Colangelo2000,Olpak2016}.
 Using the non-relativistic quark model wave functions derived from a few possible quark models \cite{Olpak2016,Melikhov2000,Ebert:2010zu,Hwang2009,Olpak2017} has been suggested as a candidate method for obtaining the leading twist LCDAs.
 This method has the benefit of calculating the LCDAs corresponding to the radially excited states.

In this study, we focus on the correlation function which involves the coupling of on-shell $1^{++}$ charmonium state to $0^{-}$ and $1^{-}$ currents.
$D$ and $D^*$ mesons that make up the molecular state will be represented by these $0^{-}$ and $1^{-}$ currents (at the hadronic level) coupling to the $1^{++}$ charmonium state.
In light-cone QCD sum rules, this correlation function can be used for calculating the relevant coupling parameters.
 Calculations have been performed at twist-3 accuracy, revealing the intricate nature of charmonium states and their interactions.
 The article begins by discussing the theoretical framework of QCD sum rules for the coupling parameters.
 Following this, a detailed calculation of the correlation function is presented.
 Coupling parameters are then extracted from the results of this calculation.
 The conclusion synthesizes these findings, highlighting the interplay between molecular and charmonium components in exotic hadrons like $\chi_{c1}(3872)$.

%% file: sum_rules_part.tex
\section{QCD Sum Rules for charmonium couplings to $D$ and $D^*$}
In this section, the matrix element
\begin{align}
    \langle \bar D^0 (q) D^*(p^\prime,\eta^{(\lambda)})|J_{int}|c\bar c(q+p^\prime,\epsilon^{(\sigma)})\rangle \equiv[G_1(\epsilon^{(\sigma)}\cdot q)(\eta^{*(\lambda)}\cdot q)+G_2(p^\prime \cdot q)(\epsilon^{(\sigma)}\cdot \eta^{*(\lambda)})] 
    \label{couplings}
\end{align}
where $J_{int}$ is the interaction part of the QCD Hamiltonian, will be analyzed. Here $\epsilon$ and $\eta$ are the polarization vectors and $\sigma$ and $\lambda$ correspond to the polarization states of the charmonium and $D^*$ meson, respectively (we suppress $\sigma$ and $\lambda$ later on for brevity). 
The matrix element involves two independent coupling parameters, $G_1$ and $G_2$.
In this study, we employ light-cone QCD sum rules to calculate these coupling parameters. 

To compute the coupling parameters, we begin with the correlation function given by
\begin{align}
    F_{\nu} & =-i\int d^4 z\, e^{-i\, q\cdot z}\, \langle c\bar{c}(P,\epsilon)|T \left \lbrace \bar{c}(z)\gamma_5 u(z) \, \bar{u}(0) \gamma_{\nu} \, c(0) \exp\left[i\int d^{4}yJ_{int}(y)\right] \right\rbrace |0\rangle 
\end{align}
(for details of the calculation, see \cite{Olpak2016}). Here, $T$ represents time-ordering. $\bar{c}(z)\gamma_5 u(z)$ and $\bar{u}(0) \gamma_{\nu} \, c(0)$ correspond to the local interpolating current operators for $D$ and $D^*$ mesons, respectively. One Taylor expands the interaction operator up to the first non-zero contribution, which here reads: 
\begin{align}
    \exp\left[i\int d^{4}yJ_{int}(y)\right]\approx 1+i\int d^{4}yJ_{int}(y).
\end{align}
In order to obtain an expression for the correlation function in terms of the hadronic parameters (like mass, decay constant, etc.), one inserts a resolution of identity in terms of on-shell hadronic states: 
\begin{align}
    1&=|0\rangle \langle 0|+\sum_{h,\vec{\eta}} \int \frac{d^3k}{(2\pi)^3 2E(\vec{k})} |h(\vec{k},\vec{\eta})\rangle \langle h(\vec{k},\vec{\eta})| \nonumber \\ 
    &+\sum_{h,h',\vec{\eta},\vec{\eta}'} \int \frac{d^3k d^3k'}{(2\pi)^6 2E(\vec{k})2E(\vec{k}')} |h(\vec{k},\vec{\eta})h'(\vec{k}',\vec{\eta}')\rangle \langle h(\vec{k},\vec{\eta})h'(\vec{k}',\vec{\eta}')|+...,
\end{align}
where the summations are over hadron types $h,h'$ and polarization states $\vec{\eta},\vec{\eta}'$, and $E(\vec{k})=\sqrt{\vec{k}^2+m_h^2}$, $m_h$ being the mass of hadron $h$. 

Using \ref{couplings} and summing over the polarization states of $D^*$ meson, one obtains: 
\begin{align}
    F^*_{\nu}&=-\frac{m_{\bar{D}}^2 m_{D^*} f_{\bar{D}} f_{D^*}}{m_c (m_D^2-q^2)(m_{D^*}^2-p'^2)}
    \\\nonumber&\quad\times\left \lbrace \frac{\epsilon \cdot q\, p' \cdot q}{m_{D^*}^2}\left( G_1 - G_2\right)p'_{\nu}-G_1\, \epsilon \cdot q\, q_{\nu}
    -G_2\, p'\cdot q\, \epsilon_{\nu}\right \rbrace.
\end{align}
Here, $f_D$ and $f_{D^*}$ are the leptonic decay constants of $D$ and $D^*$ mesons and they are defined by the following matrix elements (\cite{Olpak2016}):
\begin{align}
    \langle 0|\bar{u}\gamma_{\nu}c|D^*(p',\lambda)\rangle &= f_{D^*}m_{D^*}\eta_{\nu}^{(\lambda)},\nonumber \\
    \langle 0|\bar{u}i\gamma_{5}c|D(q)\rangle &= \frac{f_D m_D^2}{m_c+m_u} \approx \frac{f_D m_D^2}{m_c}
\end{align}

In the lowest order approximation for the light-quark propagator, the correlation function reduces to (\cite{Colangelo2000}): 
\begin{align}
    F_{\nu} & =-i\int d^4 z\, e^{-i\, q\cdot z}\, \frac{z^{\mu}}{2\pi ^2\, z^4}\langle c\bar{c}(P,\epsilon)|:\bar{c}(z)\gamma_5 \gamma_{\mu} \gamma_{\nu} \, c(0):|0\rangle,
\end{align}
where $::$ represents normal ordering.

The following definitions for the matrix elements relevant to our calculations are given in \cite{Yang2007}:
\begin{align}
    \langle c\bar{c}(P,\epsilon)| \bar{c}(y)\, \sigma_{\mu \nu}\gamma_5\, c(x)|0\rangle & = f^{\perp}_{c\bar{c}}\int_0^1 du\, 
    e^{i(u\, p\cdot y+\bar{u}\, p\cdot x)}\Big \lbrace \left(\epsilon^*_{\perp \mu}\, p_{\nu}-\epsilon^*_{\perp \nu}\, p_{\mu}\right)\, \Phi_{\perp}(u) \nonumber \\ 
    & + \frac{m_{c\bar{c}}^2\, \epsilon^*\cdot z_0}{(p\cdot z_0)^2}\left(p_{\mu}\, z_{0,\nu}-p_{\nu}\, z_{0,\mu}\right)h_{\parallel}^{(t)}(u)\nonumber \\ 
    & + \frac12 \left(\epsilon^*_{\perp \mu}\, z_{0,\nu}-\epsilon^*_{\perp \nu}\, z_{0,\mu}\right)\frac{m_{c\bar{c}}^2}{(p\cdot z_0)}h_3(u) \Big \rbrace ,\\
    \langle c\bar{c}(P,\epsilon)| \bar{c}(y)\, \gamma_5\, c(x)|0\rangle & =\frac12 f^{\perp}_{c\bar{c}}m_{c\bar{c}}^2\, \epsilon^*\cdot z_0\int_0^1 du\, e^{i(u\, p\cdot y+\bar{u}\, p\cdot x)}\, h^{(p)}_{\parallel}(u), 
\end{align}
where 
\begin{align}
    p_{\mu} & =P_{\mu}-\frac{m_{c\bar{c}}^2\, z_{0,\mu}}{2P\cdot z_0},\\
    z & =y-x,\\
    \epsilon_{\parallel \mu} &=\frac{\epsilon \cdot z_0}{p\cdot z_0}\left( p_{\mu} - \frac{m_{c\bar{c}}^2\, z_{0,\mu}}{2p\cdot z_0}\right),\\
    \epsilon_{\perp \mu} &=\epsilon_{\mu}-\epsilon_{\parallel \mu},
\end{align}
and as noted in \cite{Hwang2009}: 
\begin{align}
    z_{0,\mu}=z_{\mu}-\frac{P_{\mu}}{m_{c\bar{c}}^2}\left(P\cdot z - \sqrt{(P\cdot z)^2-m_{c\bar{c}}^2\, z^2}\right) 
\end{align}
is a light-like separation ($z_0^2=0$).
The functions $\Phi_{\perp}(u)$, $h_{\parallel}^{(p,t)}(u)$ and $h_3(u)$ correspond to the twist 2, 3 and 4 light-cone distribution amplitudes corresponding to a $c\bar{c}$ state, respectively. 

At leading order in $z^2$, $p\cdot z_0=P\cdot z_0 \simeq P\cdot z$, $z\cdot z_0\simeq 0$, $\epsilon_{\perp}\cdot z\simeq 0$, $\epsilon_{\perp}\cdot P\simeq 0$.
 We also have $\epsilon \cdot z_0 = \epsilon \cdot z$, $\epsilon_{\perp} \cdot z_0= 0$ and $\epsilon_{\perp}\cdot p=0$.
Hence: 
\begin{align}
    \frac{z^{\mu}}{z^4}\langle c\bar{c}(P,\epsilon)| \bar{c}(z_0)\, \sigma_{\mu \nu}\gamma_5\, c(0)|0\rangle & \simeq f^{\perp}_{c\bar{c}}\int_0^1 du\, 
    \frac{e^{iu\, P\cdot z}}{z^4}\Big \lbrace  \frac{m_{c\bar{c}}^2\, \epsilon^*\cdot z}{P\cdot z}\, z_{\nu}\, h_{\parallel}^{(t)}(u)  \\ 
    & -P\cdot z\, \left( \epsilon^*_{\nu}-\frac{\epsilon^* \cdot z}{P\cdot z}P_{\nu} + \frac{m_{c\bar{c}}^2\, \epsilon^* \cdot z}{(P\cdot z)^2}z_{\nu} \right)\Phi_{\perp}(u)\Big \rbrace ,\nonumber\\
    \frac{z_{\nu}}{z^4}\langle c\bar{c}(P,\epsilon)| \bar{c}(y)\, \gamma_5\, c(0)|0\rangle & \simeq \frac12 f^{\perp}_{c\bar{c}}m_{c\bar{c}}^2\, \epsilon^*\cdot z\frac{z_{\nu}}{z^4}\int_0^1 du\, e^{iu\, P\cdot z}\, h^{(p)}_{\parallel}(u). 
\end{align}

The correlation function 
involves 3 independent Lorentz 
structures which can be chosen to be proportional to $q_\nu$, $p'_\nu$ and $\epsilon_\nu$.
The sum rules are obtrained from the coefficients of each of these vectors.
 Let us reorganize the terms to get some idea about how twist-3 contributions are mixed with twist-2:

\begin{align}
    \frac{z^{\mu}}{z^4}\langle c\bar{c}(P,\epsilon)| \bar{c}(z_0)\, \sigma_{\mu \nu}\gamma_5\, c(0)|0\rangle & \simeq f^{\perp}_{c\bar{c}}\int_0^1 du\, 
    \Big \lbrace \left(\epsilon^* \cdot z\, P_{\nu} - P\cdot z\, \epsilon^*_{\nu} \right)\Phi_{\perp}(u)  \\ 
    & \quad+\frac{m_{c\bar{c}}^2\, \epsilon^*\cdot z\, z_{\nu}}{P\cdot z}\, \left(h_{\parallel}^{(t)}(u)-\Phi_{\perp}(u)\right)\Big \rbrace \frac{e^{iu\, P\cdot z}}{z^4},\nonumber\\
    \frac{z_{\nu}}{z^4}\langle c\bar{c}(P,\epsilon)| \bar{c}(y)\, \gamma_5\, c(0)|0\rangle & \simeq \frac12 f^{\perp}_{c\bar{c}}m_{c\bar{c}}^2\, \epsilon^*\cdot z\frac{z_{\nu}}{z^4}\int_0^1 du\, e^{iu\, P\cdot z}\, h^{(p)}_{\parallel}(u). 
\end{align}

In addition, twist-3 distribution amplitudes can be approximately expressed in terms of the twist-2 amplitudes as follows (Wandruza-Wilczek relations \cite{Wandzura:1977qf}, as also noted in \cite{Yang2007}):
\begin{align*}
h^{(t)}_{\parallel}(u) & \approx (2u-1)\left[ \int_0^u dv \frac{\Phi_{\perp}(v)}{\bar{v}} - \int_u^1 dv \frac{\Phi_{\perp}(v)}{v}\right]\nonumber \\
h^{(p)}_{\parallel}(u) & \approx 2\left[ \bar{u}\int_0^u dv \frac{\Phi_{\perp}(v)}{\bar{v}} + u\int_u^1 dv \frac{\Phi_{\perp}(v)}{v}\right].    
\end{align*}

Using these matrix elements and the Wandruza-Wilczek relations, the correlation function takes the form:
\begin{align}
    F_{\nu}=& -i\int \, d^4 z\, e^{-iq\cdot z} \Big \lbrace f^{\perp}_{c\bar{c}}\int_0^1 du\, \frac{e^{iu\, P\cdot z}}{z^4} \Big[ -i\left(\epsilon^* \cdot z\, P_{\nu} - P\cdot z\, \epsilon^*_{\nu} \right)\Phi_{\perp}(u)\nonumber \\
    & + m_{c\bar{c}}^2\, \epsilon^*\cdot z\, z_{\nu}B(u)\Big]\Big \rbrace .
    \label{Buterm}
\end{align}
where $B(u)\equiv \left \lbrace \frac12 h^{(p)}_{\parallel}(u)- \int_0^{u}du'\left[ h_{\parallel}^{(t)}(u')-\Phi_{\perp}(u') \right] \right \rbrace$. 
For later convenience, here we also define the following functions relevant to the calculation: $h_{\parallel}^{(t)}(u)-\Phi_{\perp}(u)\equiv a(u)$ and $dA(u)/du \equiv a(u)$, implying that $A(u)=\int_0^{u}du'a(u')$. Calculations on the QCD side (to lowest order) provide the following coefficients for each independent Lorentz structure: 
\begin{align}
  &  p'_\nu \Longrightarrow -\epsilon \cdot q \, f_{c\bar{c}}^{\perp}\int_0^1 du \left \lbrace  \frac{m_{c\bar{c}}^2 u\left[ 2A(u)+h_{\parallel}^{(p)}(u) \right]}{(up'-\bar{u}q)^4} + \frac{\Phi_{\perp}(u)}{(up'-\bar{u}q)^2}\right \rbrace, \nonumber \\
  &  q_\nu \Longrightarrow \epsilon \cdot q \, f_{c\bar{c}}^{\perp}\int_0^1 du \left \lbrace  \frac{-m_{c\bar{c}}^2 \bar{u}\left[ 2A(u)+h_{\parallel}^{(p)}(u) \right]}{(up'-\bar{u}q)^4} + \frac{\Phi_{\perp}(u)}{(up'-\bar{u}q)^2}\right \rbrace, \\
  &  \epsilon_\nu \Longrightarrow -f_{c\bar{c}}^{\perp}\int_0^1 du \frac{1}{(up'-\bar{u}q)^2} \left \lbrace m_{c\bar{c}}^2 \left[ A(u)-\frac12 h_{\parallel}^{(p)}(u)\right] + \Phi_{\perp}(u)\left[ up'^2-\bar{u}q^2+(u-\bar{u})p'\cdot q\right]\right \rbrace \nonumber,
  \label{sumrules}
\end{align}

In order to obtain the sum rules, Borel transformation should be applied to the correlation function to suppress the contributions of higher states and the continuum, and we obtain the following sum rules at the end: 
\begin{align}
  &  p'_\nu \Longrightarrow G_1-G_2 = -2A\left(\frac12\right)\frac{f^{\perp}_{c\bar{c}}}{f_D f_{D^*}}\frac{m_c m_{c\bar{c}}^2 m_{D^*}}{m_D^2\, \Delta m^2}e^{\frac{2(m_{D}^2+m_{D^*}^2)-m_{c\bar{c}}^2}{4M^2}}, \nonumber \\
  &  q_\nu \Longrightarrow G_1 = -A\left(\frac12\right)\frac{f^{\perp}_{c\bar{c}}}{f_D f_{D^*}}\frac{m_c m_{c\bar{c}}^2 }{m_D^2 m_{D^*}}e^{\frac{2(m_{D}^2+m_{D^*}^2)-m_{c\bar{c}}^2}{4M^2}}, \\
  &  \epsilon_\nu \Longrightarrow G_2 = \frac{f^{\perp}_{c\bar{c}}}{f_D f_{D^*}}\frac{m_c\, e^{\frac{(m_{D}^2+m_{D^*}^2)}{2M^2}}}{m_{D}^2 m_{D^*}\Delta m^2} \int_{\frac{m_{c\bar{c}}^2}{4}}^{s_0} ds\, e^{-s/M^2} \left[2A\left(\frac12\right)m_{c\bar{c}}^2-\Phi '\left(\frac12\right)\left(s-\frac{m_{c\bar{c}}^2}{4}\right)\right]\nonumber,
  \label{sumrules}
\end{align}
where $s_0=m_{c\bar{c}}^2/4+\alpha$, and $\alpha >0$. Wandruza-Wilczek relations stated above reveal that $h^{(p)}_{\parallel}(1/2)=0$.  

The $z_{\nu}$ term in Eq.(\ref{Buterm}) requires one partial integration over $u$. 
Using the definitions of $a(u)$, $A(u)$ and $B(u)$, we obtain the following integral: 
\begin{align}
    \int_0^1\, du\, \frac{a(u)\, e^{iu P\cdot z}}{P\cdot z}=A(1)e^{iP\cdot z}-A(0)-i\int_0^1\, du\, A(u)e^{iuP\cdot z}. 
\end{align}
Using the definitions of $A(u)$ and $h_{\parallel}^{(t)}(u)$, one obtains $A(0)=A(1)=0$.
 The proof of this argument is given in Appendix A. 

While calculating the Borel transform with respect to $p'^2$ and $q^2$, it has been taken into account that $D$ and $D^*$ meson masses are close to each other, so the Borel parameters corresponding to the independent momenta $p'$ and $q$ have been set equal to each other at the end.

The sum rules obtained from the coefficients of $p'_{\nu}$ and $q_{\nu}$ lead to the following relation: 
\begin{align}
    \frac{G_1-G_2}{G_1} = \frac{2m_{D^*}^2}{\Delta m^2},
    \label{Gratio}
\end{align}
as observed in \cite{Olpak2016}.
 In \cite{Olpak2016} this result was obtained up to twist-2 accuracy. Here it is observed that the relation is also valid at twist-3 accuracy.

The next task is to determine $s_0$, to be able to use the sum rule obtained from the coefficients of $\epsilon_{\nu}$. One argument comes from the sum rule for $G_2$.
 Using this sum rule, we can do: 
\begin{align}
\frac{\partial}{\partial \tau}\ln{(LHS)} & =\frac{\partial}{\partial \tau}\ln{(RHS)},\nonumber \\
\Rightarrow \frac{\beta_\tau (\tau,\alpha)}{\beta(\tau,\alpha)} & =\frac{-(m_D^2+m_{D^*}^2)}{2},
\label{lnarg}
\end{align}
where $\tau = 1/M^2$ and $\beta(\tau,\alpha)$ is the RHS of the third line of Eq.(\ref{sumrules}) and $\alpha \equiv s_0 - \frac{m_{q\bar{q}}^2}{4}$. 
The right hand side in Eq.(\ref{lnarg}) involves experimental values of the D-meson masses.
 Choosing $\alpha$ in accordance with this equation, we can substitute this value of $\alpha$ in the $G_2$-sum-rule to calculate $G_2$.
 Masses and decay constants of charmonia and bottomonia used in these calculations are presented in Table \ref{quark_model_1}. 

\begin{table}[]
    \centering
    \begin{tabular}{|c|c|c|c|c|}
    \hline
    Charmonia & Mass $(GeV)$ &  Mass $(GeV)$ & D. C. $(GeV)$ & D. C. $(GeV)$  \\
    \hline
     State & $\Lambda = \infty $ & $\Lambda = m_c^{Q.M.}$ & $\Lambda = \infty $ & $\Lambda = m_c^{Q.M.}$ \\
     \hline
     $n=1$ & 3.54 & 3.54 & 0.0959 & 0.0875 \\
     \hline
     $n=2$ & 3.97 & 3.97 & 0.0881 & 0.0741 \\
     \hline
     $n=3$ & 4.33 & 4.33 & 0.0824 & 0.0615 \\
     \hline
      Bottomonia & Mass $(GeV)$ &  Mass $(GeV)$ & D. C. $(GeV)$ & D. C. $(GeV)$  \\
    \hline
     State & $\Lambda = \infty $ & $\Lambda = m_b^{Q.M.}$ & $\Lambda = \infty $ & $\Lambda = m_b^{Q.M.}$ \\
     \hline
     $n=1$ & 9.89 & 9.89 & 0.0802 & 0.0794 \\
     \hline
     $n=2$ & 10.3 & 10.3 & 0.0832 & 0.0822 \\
     \hline
     $n=3$ & 10.6 & 10.6 & 0.0834 & 0.0820 \\
     \hline
    \end{tabular}
    \caption{Masses of charmonia calculated using quark model in \cite{Olpak2016, Olpak2017}. D.C.: Decay constant. $\Lambda$: Cut-off value for integrals over transverse momenta. $m_{c,b}^{Q.M.}$: Charm/Bottom quark mass used in quark model. $m_c^{Q.M.}= 1.628\, GeV$, $m_b^{Q.M.}= 4.977\, GeV$.}
    \label{quark_model_1}
\end{table}

Since the sum rules are approximate equations, we expect that there will be some discrepancy between the left and right hand sides of Eq. (\ref{lnarg}).
 We define a ``residual" $R$ out of this as: 
\begin{align}
   R = \frac{-\frac{\beta_\tau(\tau,\alpha)}{\beta(\tau,\alpha)} - \frac{(m_D^2+m_{D^*}^2)}{2}}{\frac{(m_D^2+m_{D^*}^2)}{2}},
\end{align}
and search for values of $\alpha$ for which this residual is not larger than a prescribed percentage, which we choose to be $30 \%$. 

 In Appendix B, we present plots for the residual plotted as a function of $\tau =1/M^2$, for various values of $\alpha = s_0 - \frac{m_{q\bar{q}}^2}{4}$.
 These plots reveal that the residuals are within the accepted range for $0.1\, GeV^2 < \alpha < 1.0\, GeV^2 $, with the exception of $n=3$ charmonium state.
 Taking this exception into account, we set the largest possible value of $\alpha $ in the given interval for which the residual is not larger than $30\, \%$, and present the dependence of the couplings (collectively denoted as ``$G$") on $M^2$ as well. The figures reveal that the couplings have stabilized values for $1\, GeV^2 < M^2 < 20\, GeV^2$ in the case of charmonia and for $20\, GeV^2 < M^2 < 100\, GeV^2$ in the case of bottomonia. 

We have also examined how the couplings evolve with $s_0$ within the range $\frac{m_{q\bar{q}}^2}{4}+0.1\, GeV^2 < s_0 < \frac{m_{q\bar{q}}^2}{4}+1.0\, GeV^2 $.
 The results are given in Figure \ref{GvsA}, at $M^2=5\, GeV^2$ for charmonia and at $M^2=25\, GeV^2$ for bottomonia.
 Although there is some variation in $G_2$ for charmonia with varying $s_0$, the corresponding effect for bottomonia appears to be negligible.

%% file: q2_correction.tex
\section{$O(\vec{q}^2)$ correction and numerical values for the coupling parameters}
One can also ask the effect of the relative momentum of $D$ and $D^*$ mesons in the bound state on the coupling parameters.
To address this question, we estimate the $O(\vec{q}^{\, 2})$ correction. 
Since we are interested in the content of $\chi_{c1}(3872)$, we can start from mass relations among $D$, $D^*$ and $\chi_{c1}(3872)$ mesons. First note that: 
\begin{align}
    \frac{\vec{q}^{\, 2}}{2\mu}\approx m_D+m_{D^*}-m_{\chi_{c1}}\Rightarrow \vec{q}^{\, 2}\approx -0.0504\, GeV^2,
\end{align}
where
\begin{align}
    \frac{1}{\mu}=\frac{1}{m_D}+\frac{1}{m_{D^*}},
\end{align}
with $m_D=1.87\, GeV$, $m_{D^*}=2.01\, GeV$ and $m_{\chi_{c1}}=3.906\, GeV$ \cite{Cincioglu2016}. 

The coupling is expressed as: 
\begin{align}
    g=\langle D(q)D^*(p',\eta)|J_{int}|c\bar{c}(P,\epsilon) \rangle =G_1\, \epsilon \cdot q\, \eta \cdot q + G_2\, \epsilon \cdot \eta \, p'\cdot q,
\end{align}
where $P=p'+q$.


Assuming that the $DD^*$ molecular state has zero orbital angular momentum (so its wavefunction is spherically symmetric), $\langle \vec{\epsilon}\cdot \vec{q}\, \vec{\eta}\cdot \vec{q}\rangle \sim \langle \vec{\epsilon}\cdot \vec{\eta}\, \vec{q}^{\, 2}/3 \rangle $ for the averages over the wavefunction.
 Noting that $E,\omega$ are slightly different than $m_D,m_{D^*}$ and $|\vec{q}^{\, 2}|$ is much smaller than $m_D,m_{D^*}$, we estimate:
\begin{align}
    g\approx -\vec{\epsilon}\cdot \vec{\eta}\left(G_1 \frac{m_{c\bar{c}}}{m_{D^*}}\frac{\vec{q}^{\, 2}}{3}+G_2 (m_D m_{D^*}+\vec{q}^{\, 2})\right). 
\end{align}
$\vec{\epsilon}\cdot \vec{\eta}$ is a polarization overlap and inside ${\chi_{c1}}$ meson $\epsilon$ and $\eta$ are basically the same, so:
\begin{align}
    g\sim -\left(G_1 \frac{m_{c\bar{c}}}{m_{D^*}}\frac{\vec{q}^{\, 2}}{3}+G_2 (m_D m_{D^*}+\vec{q}^{\, 2})\right). 
\end{align}
We take the values from the calculation involving twist-2 and twist-3 contributions, and take the Borel mass $M^2=5\, GeV^2$ and $s_0 = \frac{m_{c\bar{c}}^2}{4} + 1.0\, GeV^2$.
 Switching to the normalization in \cite{Cincioglu2016}, we have (see Eq.(29) and Eq.(34) in \cite{Cincioglu2016}):
\begin{align}
    d=\frac{g}{\sqrt{m_{c\bar{c}} m_D m_{D^*}}},
\end{align}
and $1\, GeV^{-1}\approx 0.2\, fm$. With the relevant values of $m_{c\bar{c}},\, G_1,\, G_2$ for $n=1,2$, Table \ref{GatOrderQ2} presents the values that we obtained. 
\begin{table}[h!]
    \centering
    \begin{tabular}{|c|c|c|c|c|}
    \hline
                                    & $\Lambda=\infty$          & $\Lambda=\infty$          & $\Lambda=m_c^{Q.M.}$      & $\Lambda=m_c^{Q.M.}$ \\
                 
                                    & $n=1$                     & $n=2$                     & $n=1$                     & $n=2$ \\
    \hline
      $m_{c\bar{c}}\, (GeV)$        & 3.54                      & 3.97                      & 3.54                      & 3.97   \\
    \hline
     $f_{c\bar{c}}\, (GeV)$         & 0.0959                    & 0.0881                    & 0.0875                    & 0.0741 \\
    \hline
      $G_1\, (GeV^{-1})$            & -5.15                     & -2.81                     & -4.75                     & -2.12  \\
    \hline   
      $G_2\, (GeV^{-1})$            & -1.81                     & 0.458                     & -1.69                     & 0.587  \\
    \hline
      $G_1 - G_2\, (GeV^{-1})$      & -3.34                     & -3.27                     & -3.06                     & -2.70  \\
    \hline
      $\Delta G\, (GeV^{-1})$       & -8.33                     & -2.76                     & -7.68                     & -2.08  \\
    \hline
      $g\, (GeV)\, (O(1))$          & 6.79                      & -1.72                     & 6.35                      & -2.21  \\
    \hline
      $g\, (GeV)\, (O(\Vec{q}^2))$  & 6.55                      & -1.79                     & 6.13                      & -2.25  \\
    \hline
      $d\, (GeV^{-1/2})\, (O(1))$   & 1.86                      & -0.446                    & 1.74                      & -0.572 \\
    \hline   
 $d\, (GeV^{-1/2})\, (O(\Vec{q}^2))$& 1.79                      & -0.464                    & 1.68                      & -0.582 \\
    \hline
      $d\, (fm^{1/2})\, (O(1))$     & 0.832                     & -0.199                    & 0.779                     & -0.256 \\
    \hline  
 $d\, (fm^{1/2})\, (O(\Vec{q}^2))$  & 0.802                     & -0.207                    & 0.751                     & -0.260 \\
    \hline
    \end{tabular}
    \caption{Values of the $c\bar{c}\rightarrow DD^{*}$ coupling from QCD sum rules and the corresponding values for the effective theory calculations. $\Delta G=G_1-G_2$ corresponds to the ``difference of couplings" calculated from the $p'$ sum rule. $G_1-G_2$ corresponds to the difference of $G_1$ and $G_2$ calculated from $q$ and $\epsilon$ sum rules.}
    \label{GatOrderQ2}
\end{table}

A similar calculation can be carried out for $b\bar{b}\rightarrow BB^{*}$ couplings as well.
 Although we do not have another result from the literature for comparison, it is still valuable to state our results and make some comments on them.
 Borel parameter is taken to be $M^2=25\, GeV^2$ and $\vec{q}^2 \approx -0.05\, GeV^2$ estimated for the $DD^*$ state can be accepted for the $BB^*$ system as well, since this is also a heavy meson bound state. Then we obtain the values presented in Table \ref{GatOrderQ2bb} ($B$ and $B^*$ masses are taken from \cite{Olpak2016}, and $\vec{q}^2 \approx -0.05\, GeV^2$ requires the bound state mass to be around $10.62\, GeV$, slightly smaller than the sum of $B$ and $B^*$ masses).
\begin{table}[h!]
    \centering
    \begin{tabular}{|c|c|c|c|c|}
    \hline
                                    & $\Lambda=\infty$          & $\Lambda=\infty$          & $\Lambda=m_b^{Q.M.}$      & $\Lambda=m_b^{Q.M.}$ \\
                 
                                    & $n=1$                     & $n=2$                     & $n=1$                     & $n=2$ \\
    \hline
      $m_{b\bar{b}}\, (GeV)$        & 9.89                      & 10.3                      & 9.89                      & 10.3   \\
    \hline
     $f_{b\bar{b}}\, (GeV)$         & 0.0802                    & 0.0832                    & 0.0794                    & 0.0822 \\
    \hline
      $G_1\, (GeV^{-1})$            & -15.1                     & -9.20                     & -15.0                     & -9.09  \\
    \hline   
      $G_2\, (GeV^{-1})$            & -0.27                     & 0.100                     & -0.269                    & 0.0990 \\
    \hline
      $G_1 - G_2\, (GeV^{-1})$      & -14.9                     & -9.30                     & -14.7                     & -9.19  \\
    \hline
      $\Delta G\, (GeV^{-1})$       & -20.7                     & -10.5                     & -20.5                     & -10.4  \\
    \hline
      $g\, (GeV)\, (O(1))$          & 7.64                      & -2.82                     & 7.56                      & -2.79  \\
    \hline
      $g\, (GeV)\, (O(\Vec{q}^2))$  & 7.15                      & -3.11                     & 7.08                      & -3.08  \\
    \hline
      $d\, (GeV^{-1/2})\, (O(1))$   & 0.458                     & -0.166                    & 0.453                     & -0.164 \\
    \hline   
 $d\, (GeV^{-1/2})\, (O(\Vec{q}^2))$& 0.43                      & -0.183                    & 0.425                     & -0.181 \\
    \hline
      $d\, (fm^{1/2})\, (O(1))$     & 0.205                     & -0.0740                   & 0.203                     & -0.0731\\
    \hline  
 $d\, (fm^{1/2})\, (O(\Vec{q}^2))$  & 0.192                     & -0.0817                   & 0.190                     & -0.0807 \\
    \hline
    \end{tabular}
    \caption{Values of the $b\bar{b}\rightarrow BB^{*}$ coupling from QCD sum rules. $\Delta G=G_1-G_2$ corresponds to the ``difference of couplings" calculated from the $p'$ sum rule. $G_1-G_2$ corresponds to the difference of $G_1$ and $G_2$ calculated from $q$ and $\epsilon$ sum rules.}
    \label{GatOrderQ2bb}
\end{table}

Let us now interpret these results for estimating the charmonium content of $\chi_{c1}(3872)$.
 In \cite{Cincioglu2016}, the parameter $d$ is shown to be proportional to the overlap of charmonium and $DD^*$ molecule states.
 Owing to this, the results of this work (within QCD sum rules) are compared to the findings of \cite{Cincioglu2016} (within effective field theory), which also includes direct interpretation of the coupling in terms of its relation to corresponding charmonium content of $\chi_{c1}(3872)$.
 Calculations made in this work for the parameter $d$ do not distinguish between the polarization states of the charmonium and the $DD^*$ molecule, so we compare our results to the results presented in \cite{Cincioglu2016} in terms of absolute values. 

From Table \ref{GatOrderQ2}, it is observed that the coupling parameter for $n=2$ (1st radially excited charmonium state) is around $|d|=0.20-0.21 fm^{1/2}$ for calculations assuming no cut-off in the calculation of decay constants and distribution amplitudes ($\Lambda = \infty$), and around $|d|=0.25-0.26 fm^{1/2}$ for calculations assuming a cut-off $\Lambda = m_c^{Q.M.}$ in the calculation of decay constants and distribution amplitudes.
 Table 1 of \cite{Cincioglu2016} states that increasing $|d|$ implies decreasing molecular content, hence increasing charmonium content. According to Table \ref{GatOrderQ2}, charmonium content of $\chi_{c1}(3872)$ is around 25-35\%, according to our calculations. Also, it should be noted that $O(\vec{q}^2)$ correction to $|d|$ does not appear to be significant within the accuracy achieved in this study. 

\begin{figure}
    \centering
    \includegraphics[width=0.6\linewidth]{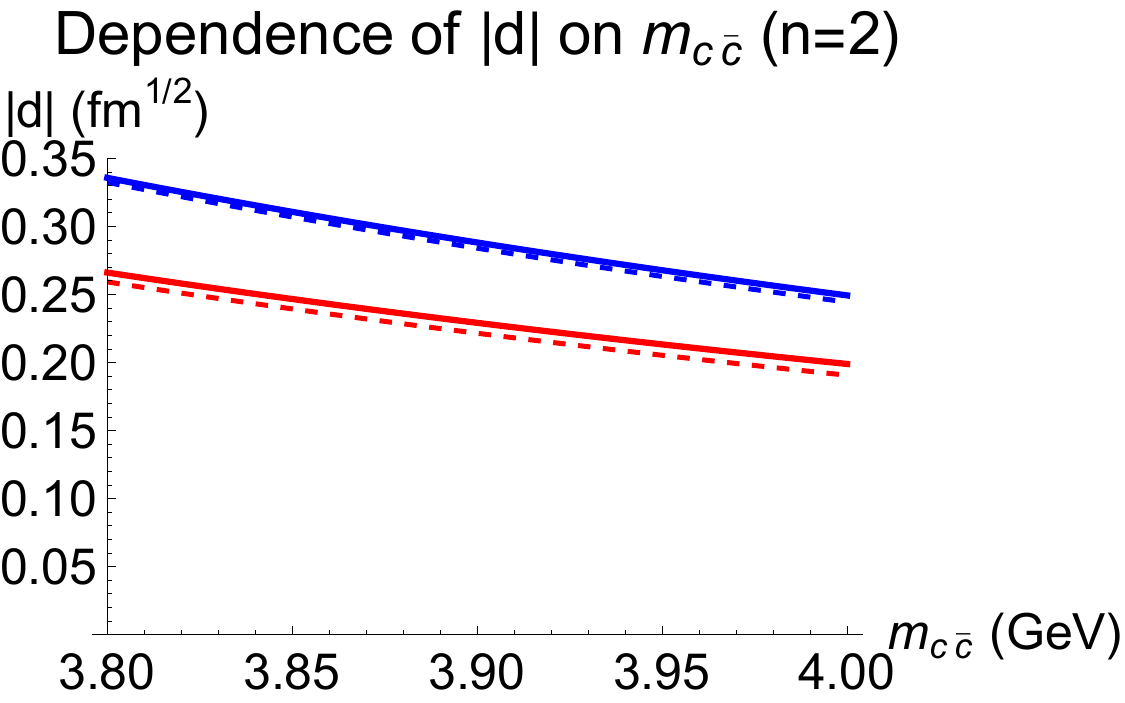}
    \caption{Dependence of the coupling paremeter $|d|$ on $m_{c\bar{c}}$. Red, dashed: $\Lambda=\infty$, $O(1)$. Red, thick: $\Lambda=\infty$, $O(\vec{q}^2)$. Blue, dashed: $\Lambda=m_{c\bar{c}}^{Q.M.}$, $O(1)$. Red, thick: $\Lambda=m_{c\bar{c}}^{Q.M.}$, $O(\vec{q}^2)$.}
    \label{dvsmcc}
\end{figure}

An important parameter in calculating the coupling parameter $d$ is the mass of the relevant charmonium state.
 In Appendix B, we present plots of residuals calculated for various values of $n=2$ charmonium mass ($m_{c\bar{c}}$), and show that the residuals are still within our prescribed uncertainty range (below 30\%) (see Figure \ref{residualsnew}).
 Relying on this observation, we present in Figure \ref{dvsmcc} how $|d|$ depends on $m_{c\bar{c}}$ within the range from $3.8\, GeV$ up to $4.0\, GeV$.
 Since lower $|d|$ implies lower charmonium contribution, we understand from the plot that the charmonium contribution to $\chi_{c1}(3872)$ is at least around 25\% (according to Table 1 of \cite{Cincioglu2016}). 

The results for the $d$ parameter calculated for charm quark and bottom quark systems suggest that $\chi_{b1}(nP)$ spectrum (with $n$ being the radial quantum number) might include a molecule candidate, which might also possess a measurable bottomonium contribution.
 This conclusion can be drawn with the assumption that the molecule and quarkonium contributions scale roughly in the same manner as function of the $d$ parameter for both charm quark and bottom quark systems. 
 However, according to Particle Data Group (PDG) listings \cite{ParticleDataGroup:2024cfk}, $\chi_{b1}(nP)$ spectrum contains the following masses: $\chi_{b1}(1P)$ mass is $9.892\, GeV$; $\chi_{b1}(2P)$ mass is $10.255\, GeV$, which is slightly smaller than the calculated bottomonium mass of $10.3\, GeV$ presented in this work; and $\chi_{b1}(3P)$ mass is $10.513\, GeV$, which is close to the sum of $B$ and $B^*$ masses (which equals to $10.65\, GeV$ (\cite{ParticleDataGroup:2024cfk})). However, in our calculations, the results are presented for $n=1,2$ states, which suggests that a more detailed analysis is needed to give a conclusive result for an exotic $\chi_{b1}$ state (if it exists). 

%% file: conclusion.tex
\section{Conclusion}

In this work, charmonium content of $\chi_{c1}(3872)$ has been studied within the context of light-cone sum rules, at twist 3 accuracy. It has been shown that, at this accuracy, charmonium content of $\chi_{c1}(3872)$ cannot be smaller than 25\%. Owing to this result, it can be concluded that $\chi_{c1}(3872)$ is neither a pure molecular state, nor a pure quarkonium state. Comparable contributions from the quarkonium and molecular states to the content of $\chi_{c1}(3872)$ also imply that the tetraquark contribution to $\chi_{c1}(3872)$ is also established, providing further evidence on the existence of exotic hadron states. The fact that our analysis is purely theoretical and the only input parameter related to $\chi_{c1}(3872)$ is the mass parameter used in the effective field theory description of \cite{Cincioglu2016} strengthens this conclusion. In addition to these, it is also argued that $\chi_{b1}$ spectrum might also contain a candidate exotic hadron, but this point requires a more detailed analysis.

%% file: appendix.tex
\section{Appendix A}
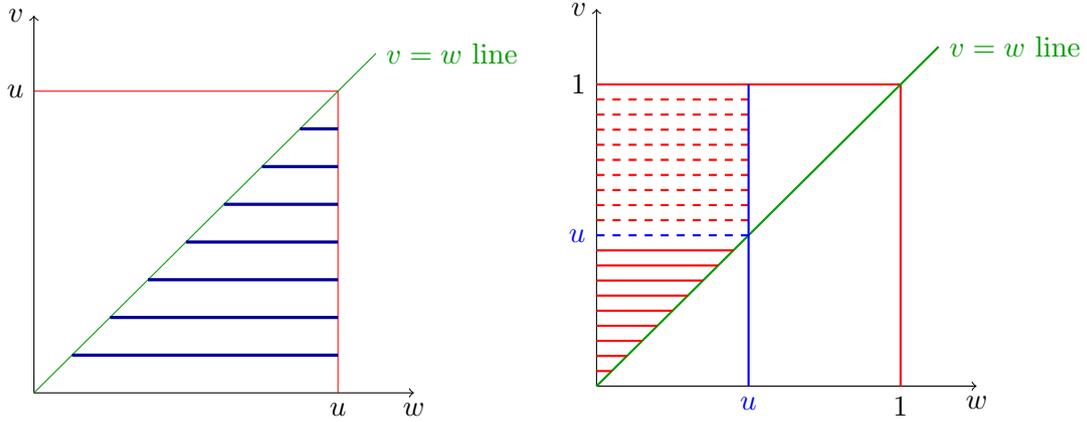
\begin{figure}[h]
\begin{tabular}{cc}
\begin{tikzpicture}

\draw[->] (0,0) -- (5,0) node[below] {$w$};
\draw[->] (0,0) -- (0,5) node[left] {$v$};

\draw[red] (0,4) node[left] {$\textcolor{black}{u}$} -- (4,4) -- (4,0);

\draw[green!60!black] (0,0) -- (4.5,4.5) node[right] {$v=w$ line};

\foreach \i in {0.5,1,1.5,2,2.5,3,3.5} {
    \draw[very thick,blue!60!black] (4,\i) -- (\i,\i);
}

\node[below] at (4,0) {$u$};

\end{tikzpicture}
&
\begin{tikzpicture}

\draw[->] (0,0) -- (5,0) node[below] {$w$};
\draw[->] (0,0) -- (0,5) node[left] {$v$};

\draw[red,thick] (0,4) node[left] {\textcolor{black}{1}} -- (4,4) -- (4,0) node[below] {\textcolor{black}{1}};

\draw[green!60!black,thick] (0,0) -- (4.5,4.5) node[right] {$v=w$ line };

\foreach \i in {0.2,0.4,0.6,0.8,1,1.2,1.4,1.6,1.8} {
    \draw[red,thick] (0,\i) -- (\i,\i);
}

\foreach \i in {2.2,2.4,2.6,2.8,3,3.2,3.4,3.6,3.8} {
    \draw[red,thick,dashed] (0,\i) -- (2,\i);
}

\draw[blue,thick] (2,0) node[below] {$u$} -- (2,2);
\draw[blue,thick,dashed] (0,2) node[left] {$u$} -- (2,2);
\draw[blue,thick] (2,2) -- (2,4);

\end{tikzpicture}
\end{tabular}
\caption{Regions of integration for the second integral is shown on the left. Region of integration for the third integral is shown on the right.}
\label{ROI}
\end{figure}

The proof for $A(0)=A(1)=0$ is as follows (notice that in \cite{Yang2007}, $\xi =2u-1$): 
\begin{align}
    A(u) & =\int_0^u dw\left(h_{\parallel}^{(t)}(w)-\Phi_{\perp}(w)\right)\nonumber \\
         & = \int_0^u dw\left \lbrace -\Phi_{\perp}(w)+(2w-1)\left[\int_0^w dv \frac{\Phi_{\perp}(v)}{1-v}-\int_w^1 dv \frac{\Phi_{\perp}(v)}{v}\right] \right \rbrace .
\end{align}
For the second and third integrals, we need to change the order of integrations within the double integrals. Consider the second one (the regions of integration are shown in Figure \ref{ROI}): 
\begin{align}
    \int_0^u dw(2w-1)\int_0^w dv \frac{\Phi_{\perp}(v)}{1-v} & = \int_0^u dv\int_v^u dw (2w-1) \frac{\Phi_{\perp}(v)}{1-v}\nonumber \\
                                                             & = \int_0^u dv \frac{\Phi_{\perp}(v)}{\bar{v}}(-u\bar{u}+v\bar{v}). 
\end{align}


For the third integral, one obtains:
\begin{align}
&    \int_0^u dw(2w-1)\int_w^1 dv \frac{\Phi_{\perp}(v)}{v}\nonumber \\
& = \int_0^u dw(2w-1)\int_w^u dv \frac{\Phi_{\perp}(v)}{v}+\int_0^u dw(2w-1)\int_u^1 dv \frac{\Phi_{\perp}(v)}{v} \nonumber \\
& = \int_0^u dv \int_0^v dw (2w-1) \frac{\Phi_{\perp}(v)}{v} - u\bar{u}\int_u^1 dv \frac{\Phi_{\perp}(v)}{v} \nonumber \\
& = -\int_0^u dv \bar{v}\Phi_{\perp}(v) - u\bar{u}\int_u^1 dv \frac{\Phi_{\perp}(v)}{v}.
\end{align}
Putting them all together, one obtains: 
\begin{align}
A(u)= -u\bar{u}\left[ \int_0^u dv \frac{\Phi_{\perp}(v)}{\bar{v}} - \int_u^1 dv \frac{\Phi_{\perp}(v)}{v}\right]. 
\end{align}
So, obviously $A(0)=0$, and $A(1)=0$. 

\section{Appendix B}

In this appendix, we present the plots for the residuals and for $M^2$ dependence and $s_0$ dependence of the couplings. 

\begin{figure}[ht]  
     \subfloat{\includegraphics[width=0.35\linewidth]{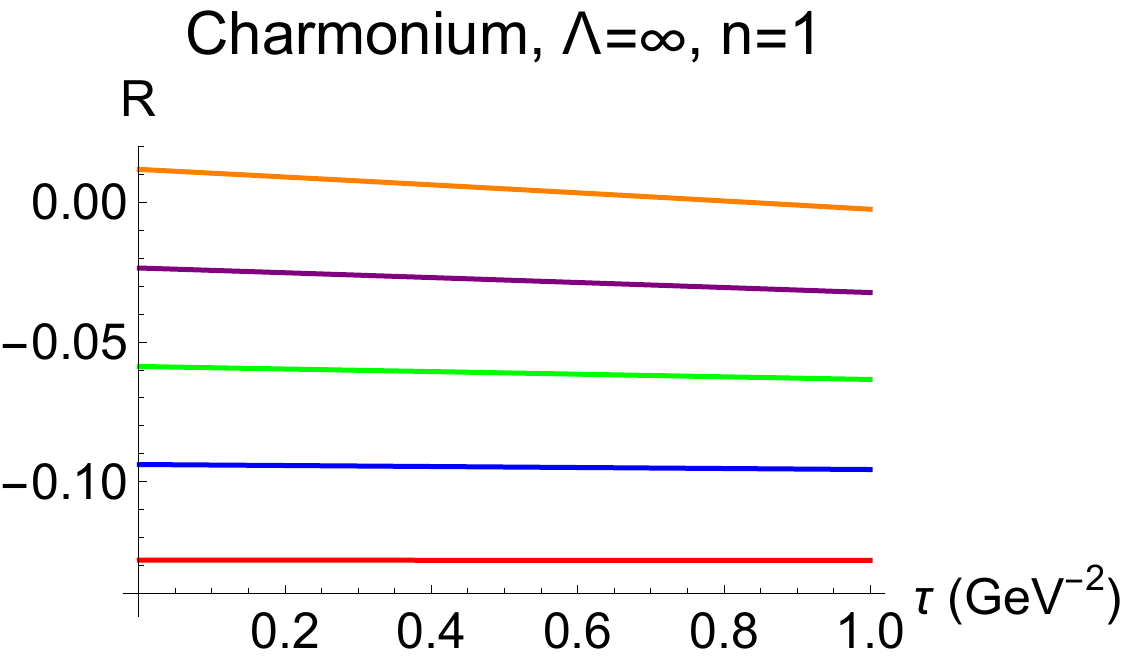}}  \subfloat{\includegraphics[width=0.35\linewidth]{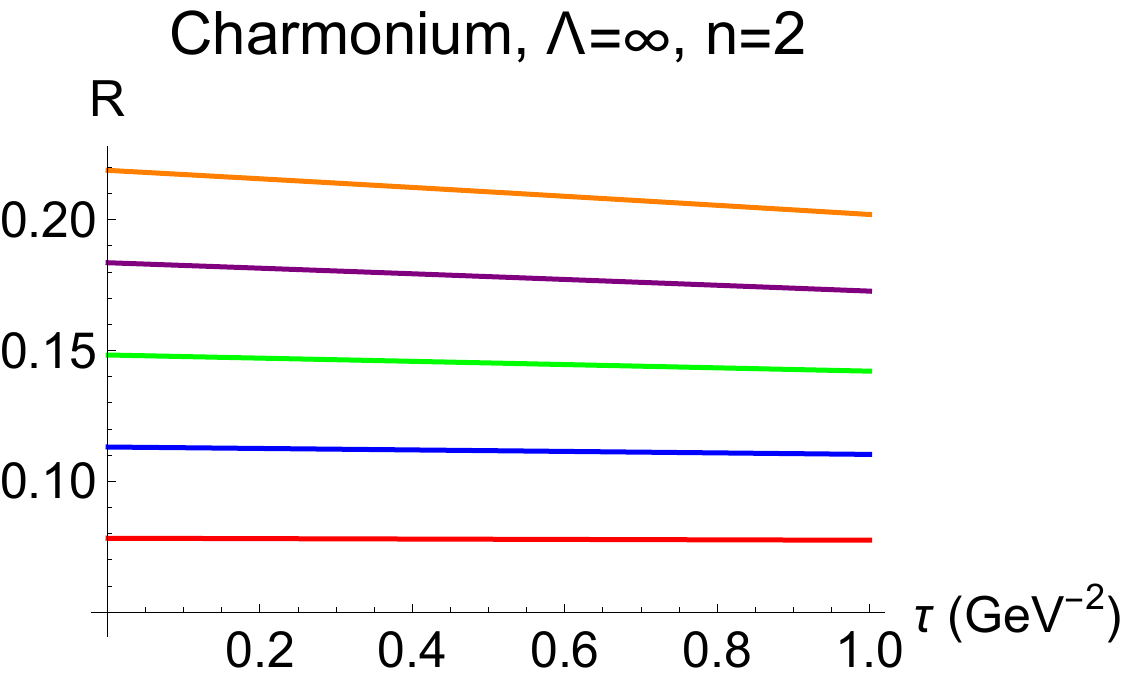}} \subfloat{\includegraphics[width=0.35\linewidth]{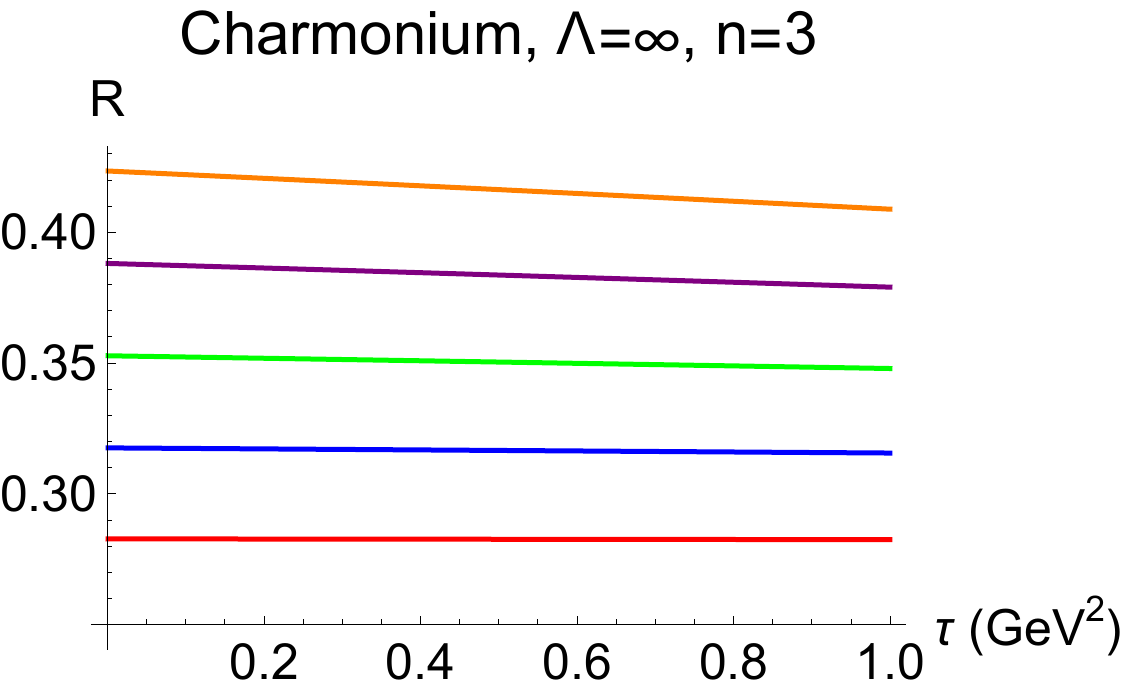}}
\\
     \subfloat{\includegraphics[width=0.35\linewidth]{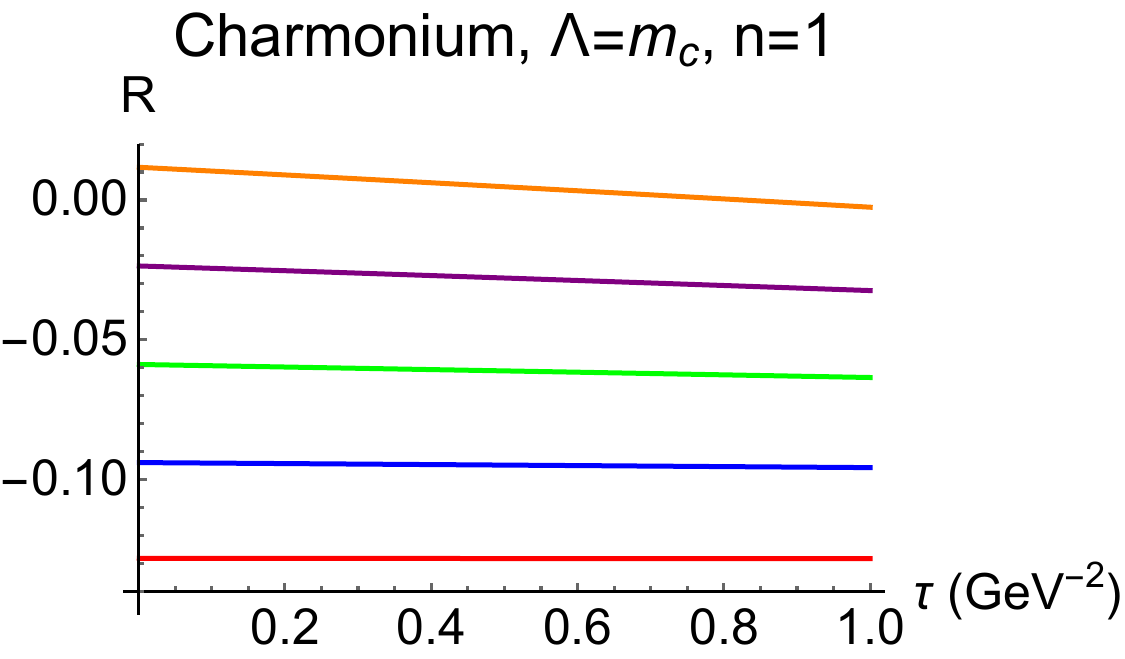}} \subfloat{\includegraphics[width=0.35\linewidth]{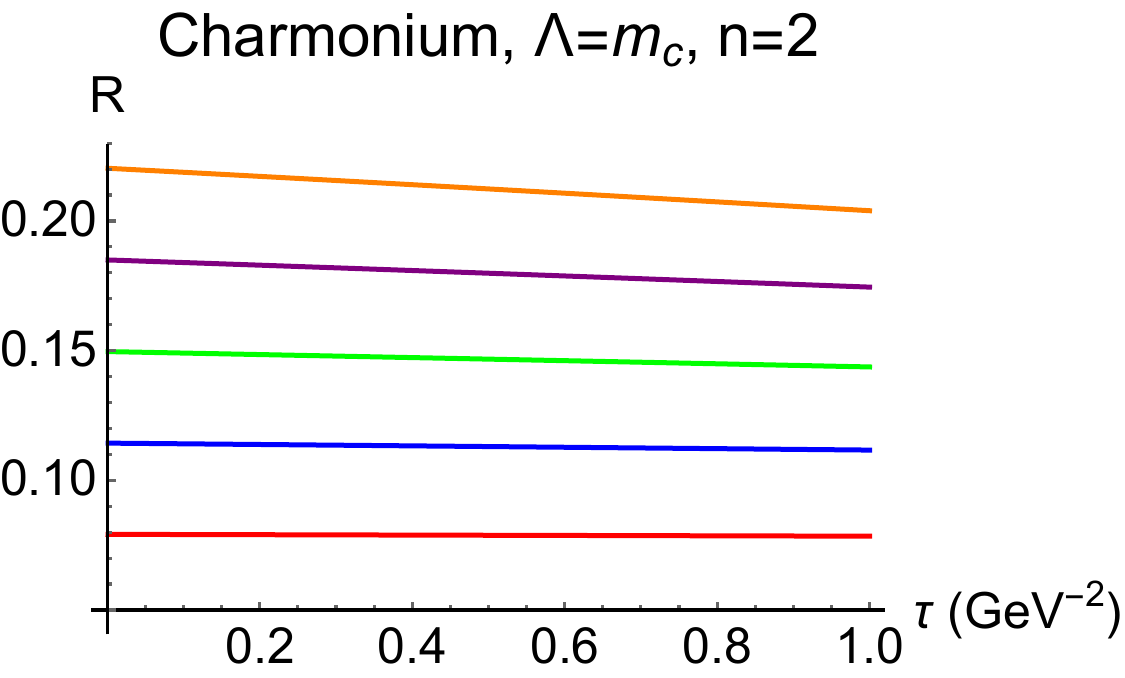}} \subfloat{\includegraphics[width=0.35\linewidth]{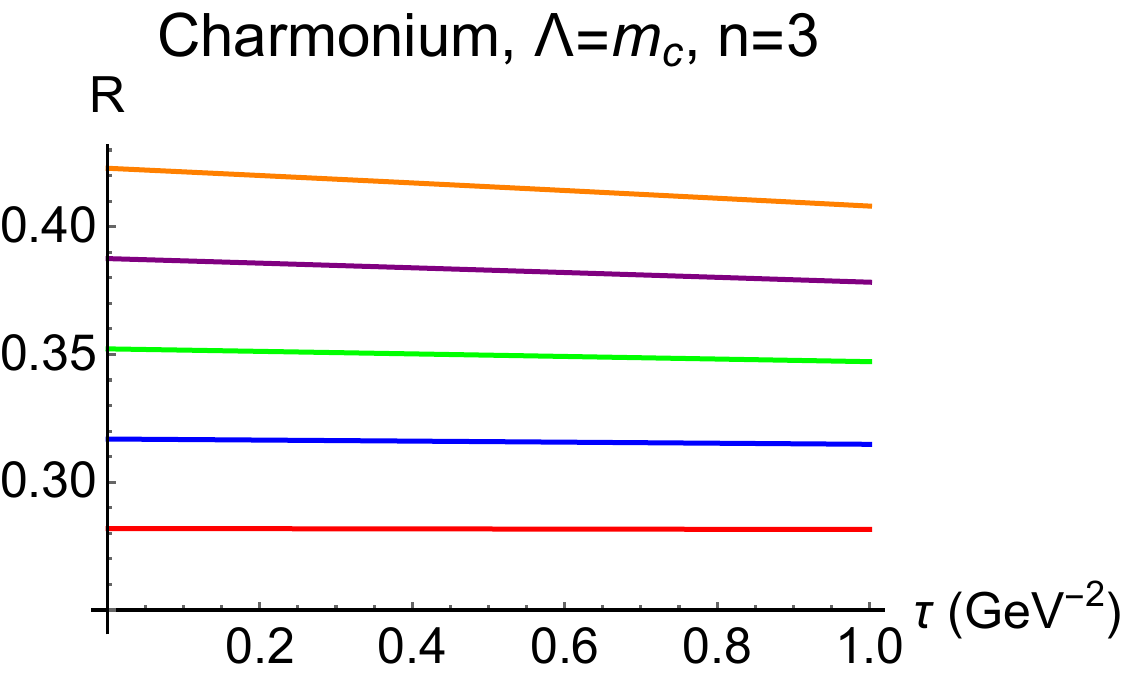}}
\\
     \subfloat{\includegraphics[width=0.35\linewidth]{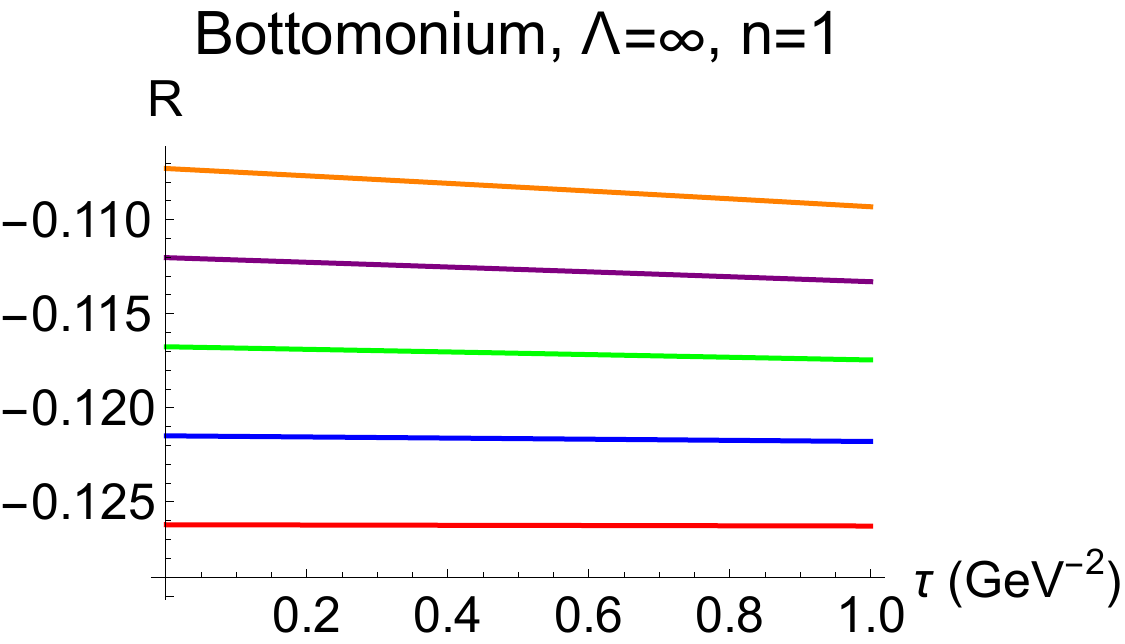}} \subfloat{\includegraphics[width=0.35\linewidth]{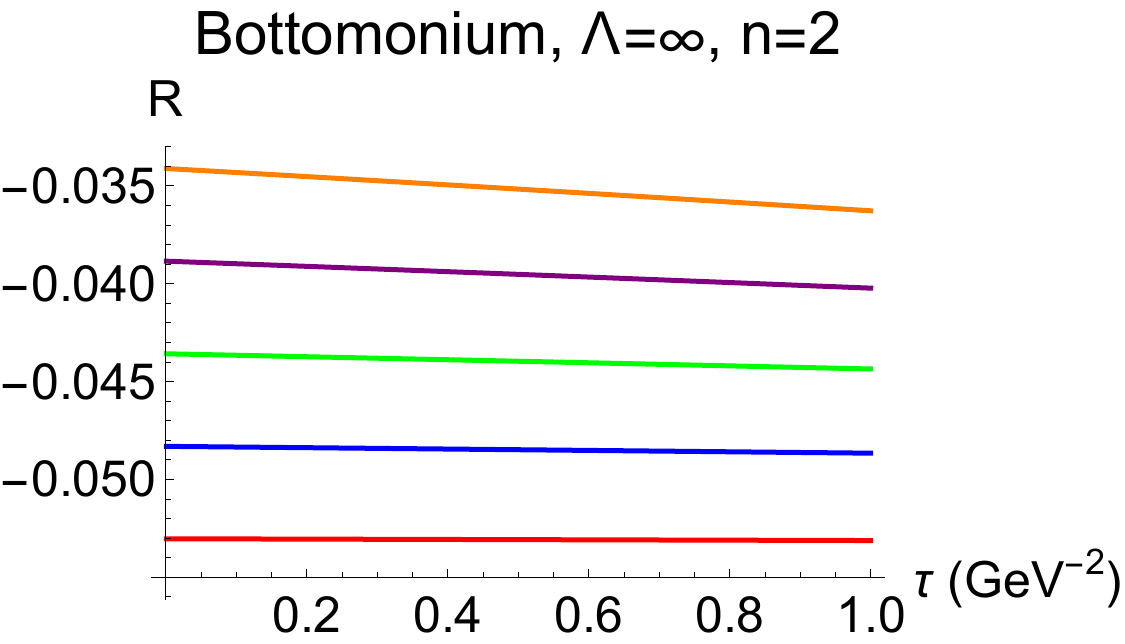}} \subfloat{\includegraphics[width=0.35\linewidth]{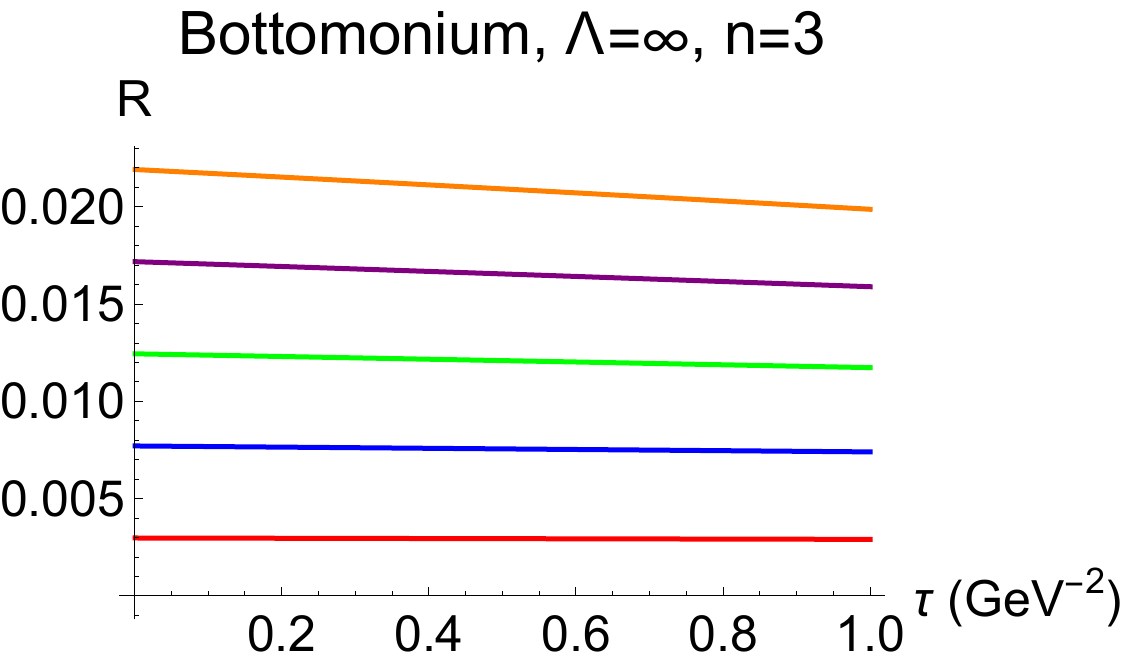}}
\\
     \subfloat{\includegraphics[width=0.35\linewidth]{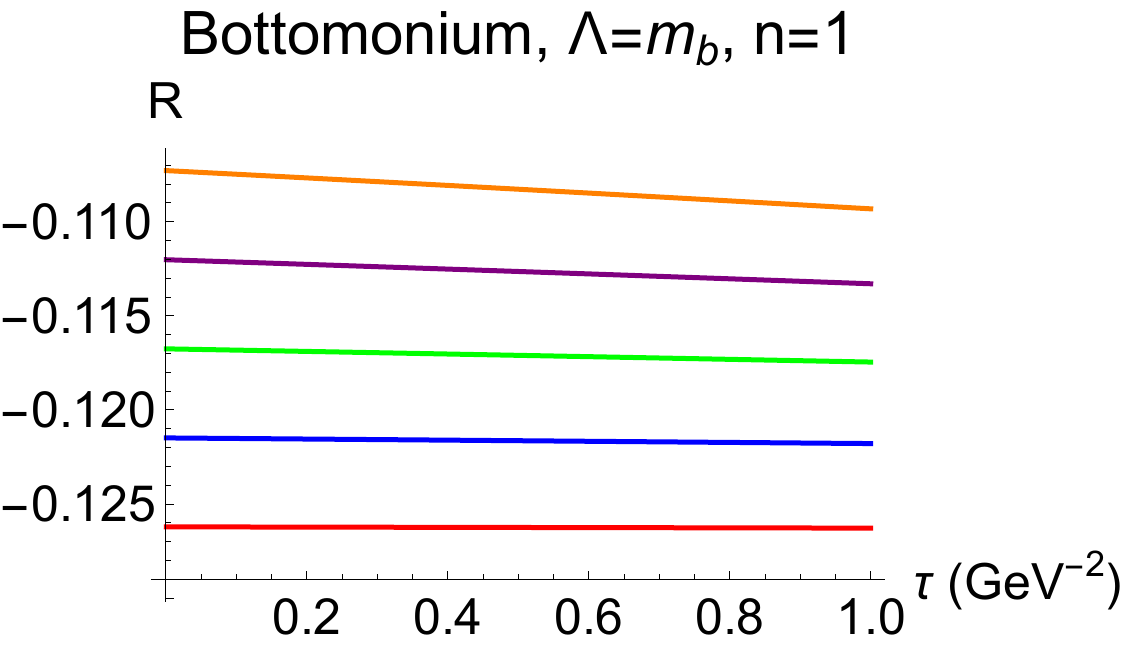}} \subfloat{\includegraphics[width=0.35\linewidth]{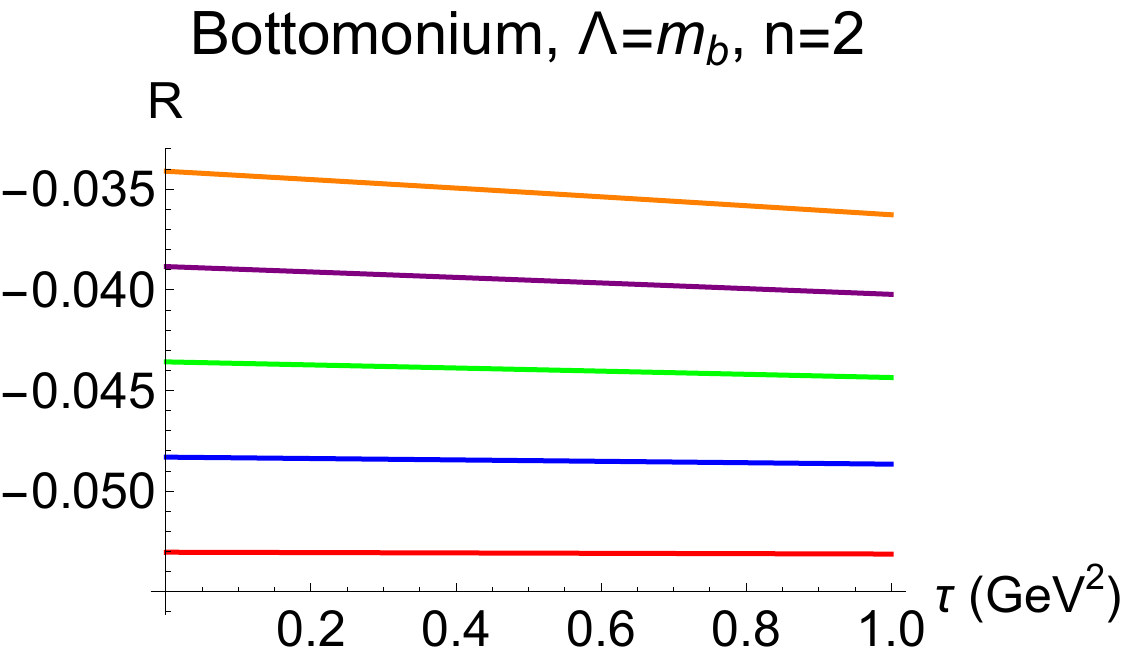}} \subfloat{\includegraphics[width=0.35\linewidth]{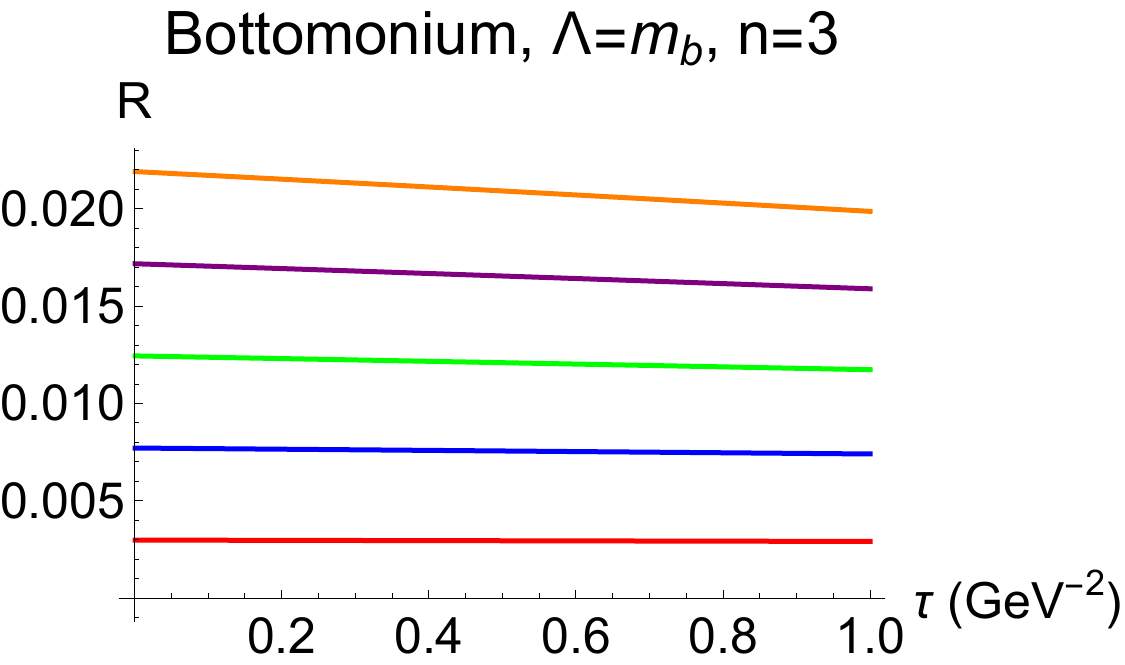}}
\\
    \caption{$R$ vs $\tau =1/M^2$. For each state, $s_0 = \frac{m_{q\bar{q}}^2}{4}+\alpha $. Red: $\alpha = 0.2\, GeV^2$. Blue: $\alpha = 0.4\, GeV^2$. Green: $\alpha = 0.6\, GeV^2$. Purple: $\alpha = 0.8\, GeV^2$. Orange: $\alpha = 1.0\, GeV^2$.}
    \label{residuals}
\end{figure}

%





\begin{figure}[ht]

     \subfloat{\includegraphics[width=0.35\linewidth]{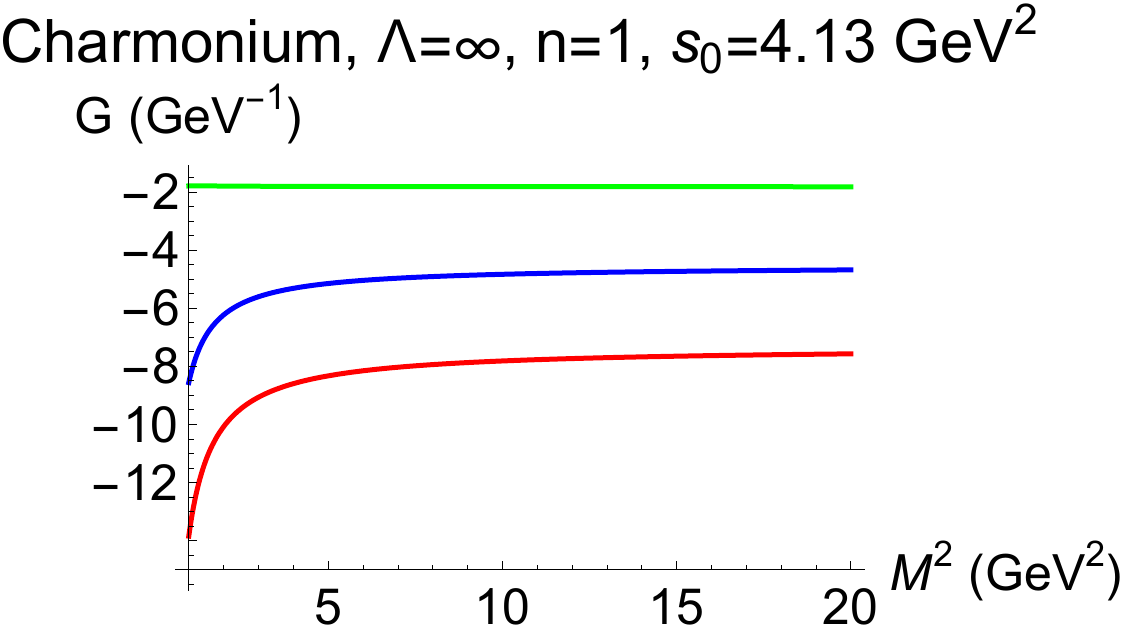}} \subfloat{\includegraphics[width=0.35\linewidth]{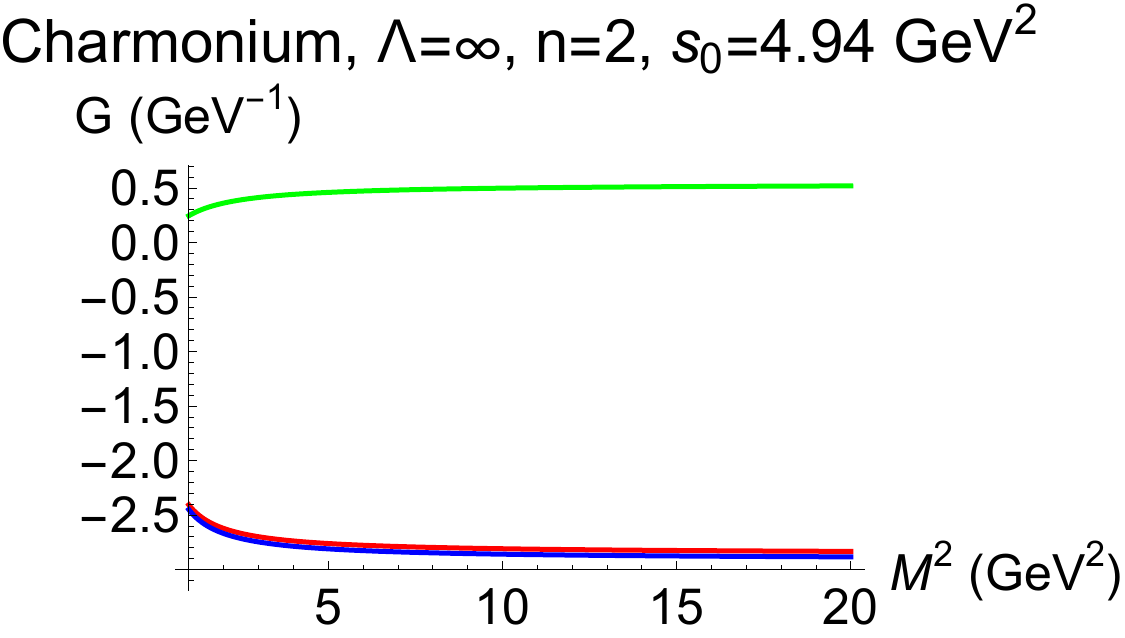}} \subfloat{\includegraphics[width=0.35\linewidth]{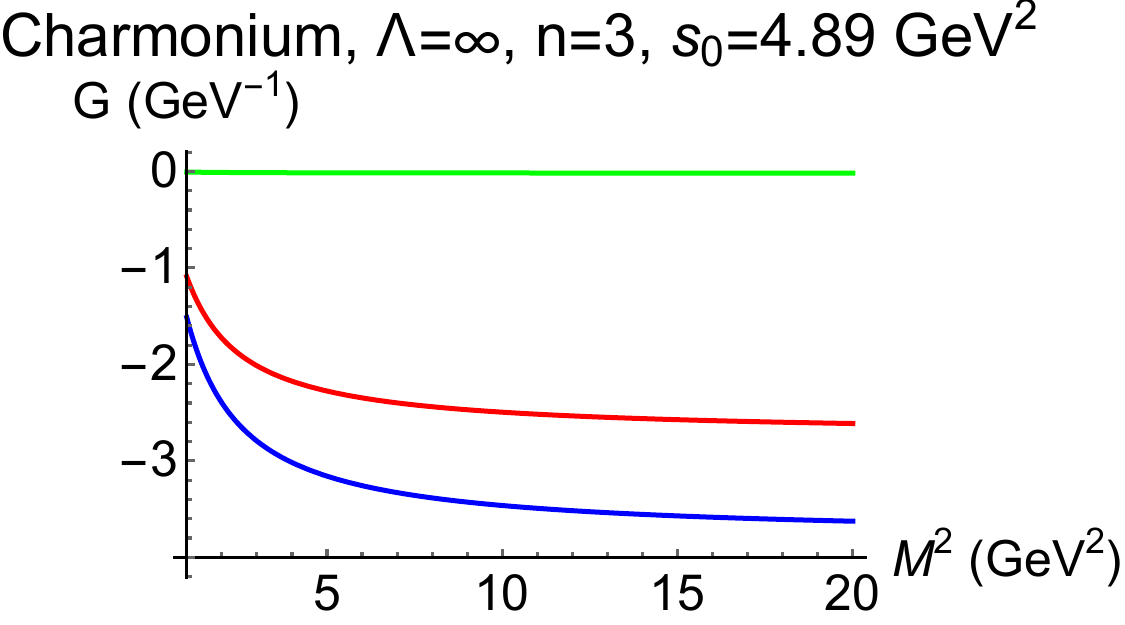}}
\\
     \subfloat{\includegraphics[width=0.35\linewidth]{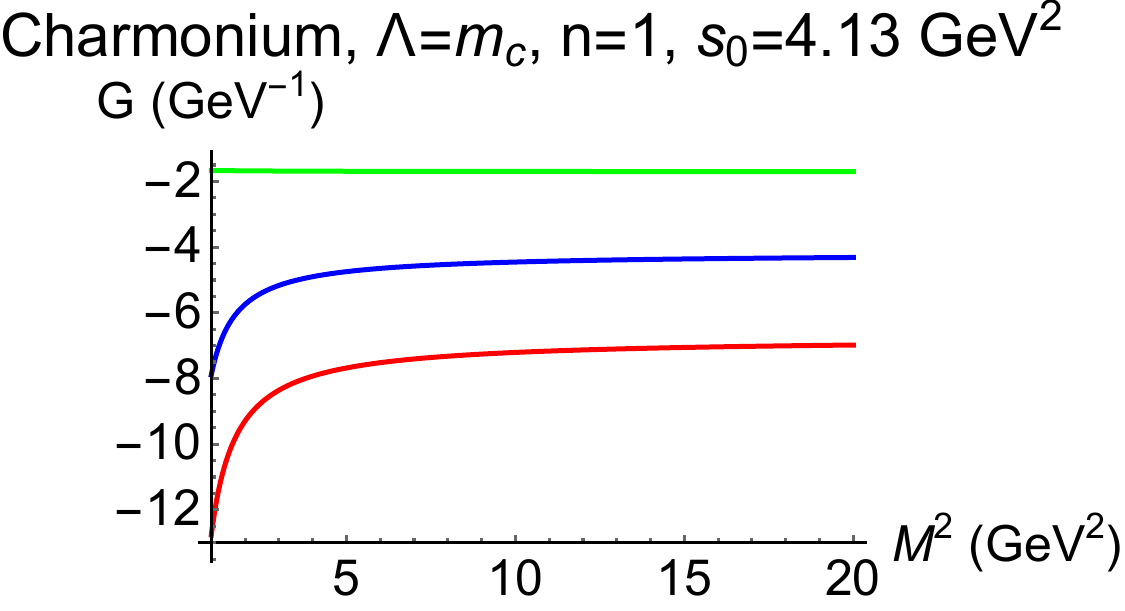}} \subfloat{\includegraphics[width=0.35\linewidth]{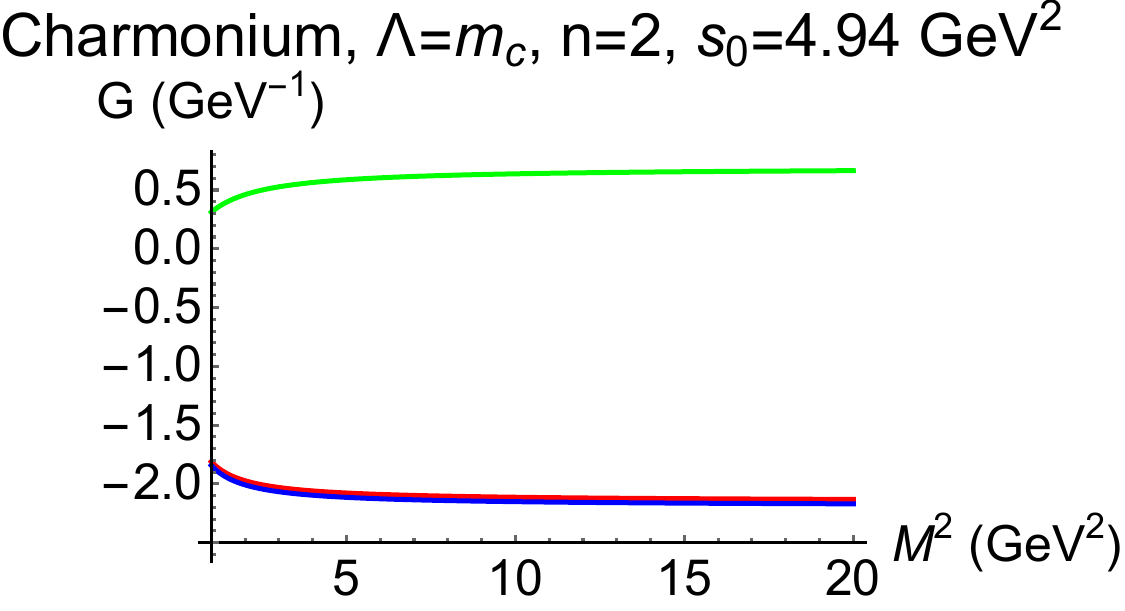}} \subfloat{\includegraphics[width=0.35\linewidth]{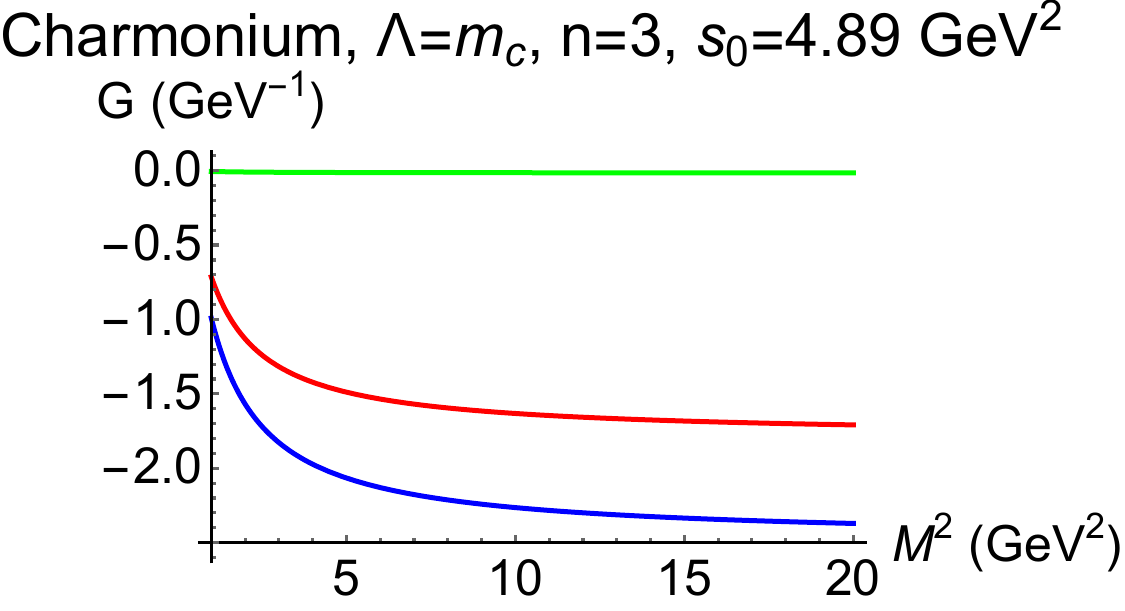}}
\\
     \subfloat{\includegraphics[width=0.35\linewidth]{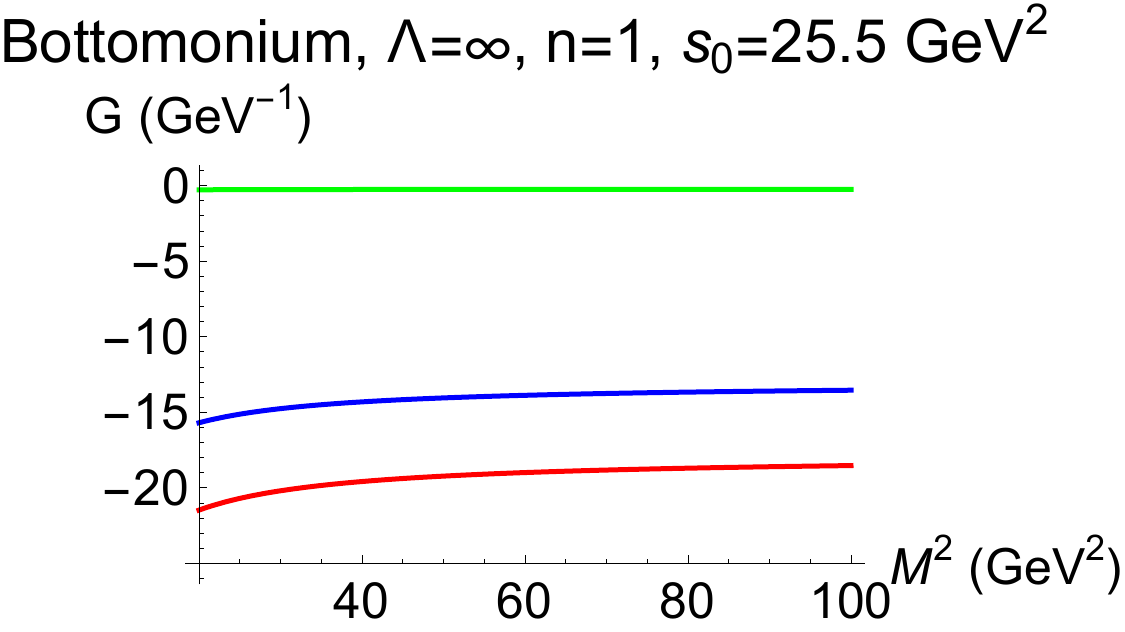}} \subfloat{\includegraphics[width=0.35\linewidth]{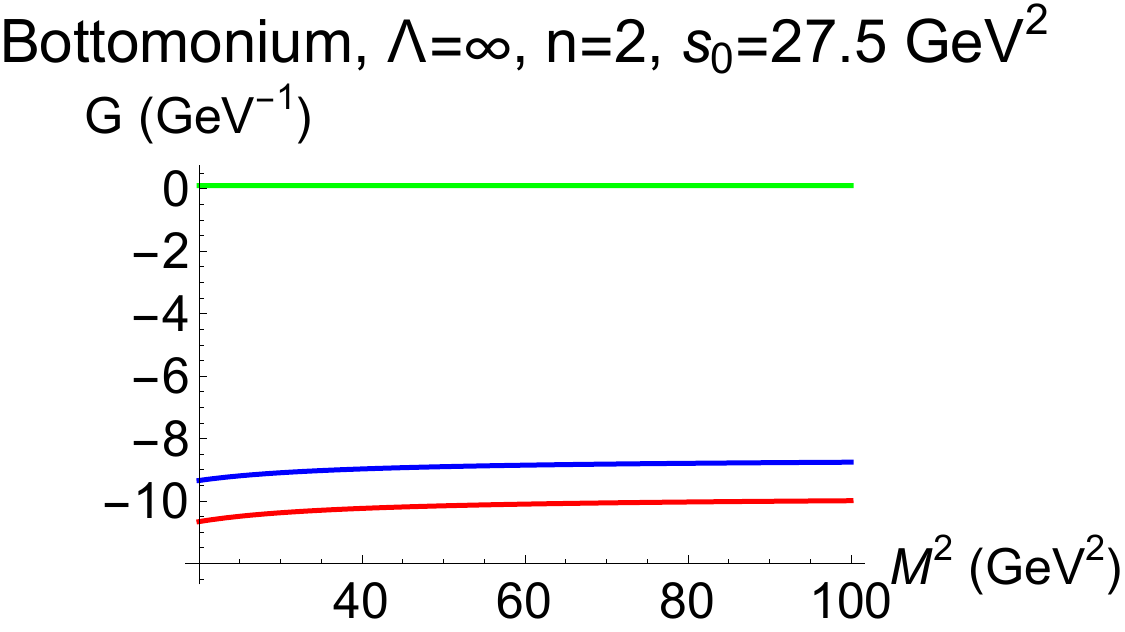}} \subfloat{\includegraphics[width=0.35\linewidth]{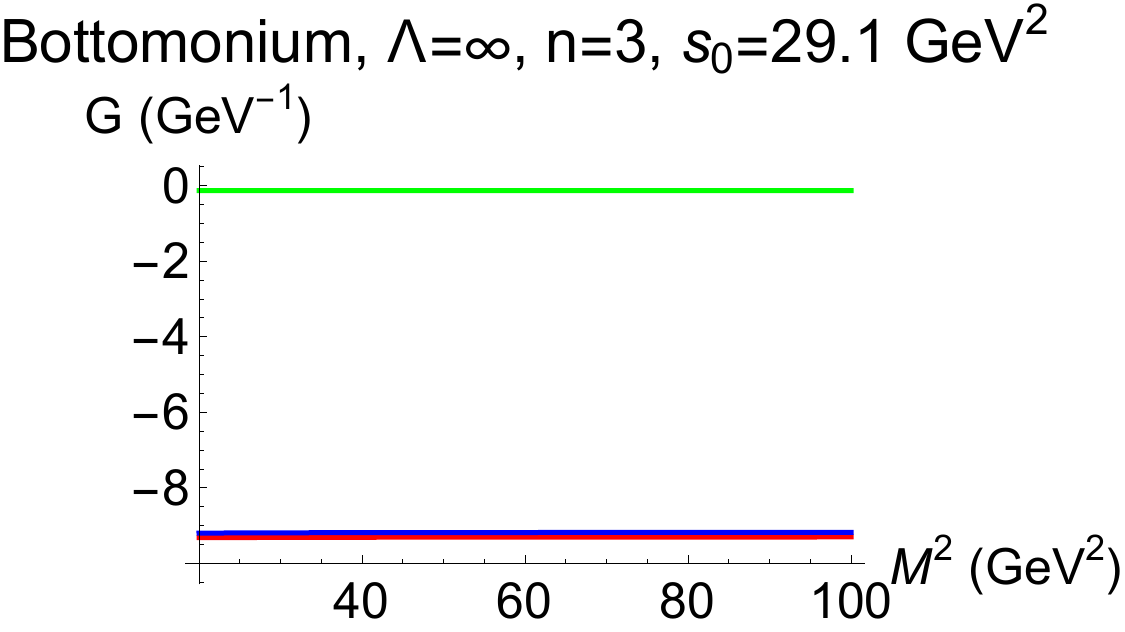}}
\\
     \subfloat{\includegraphics[width=0.35\linewidth]{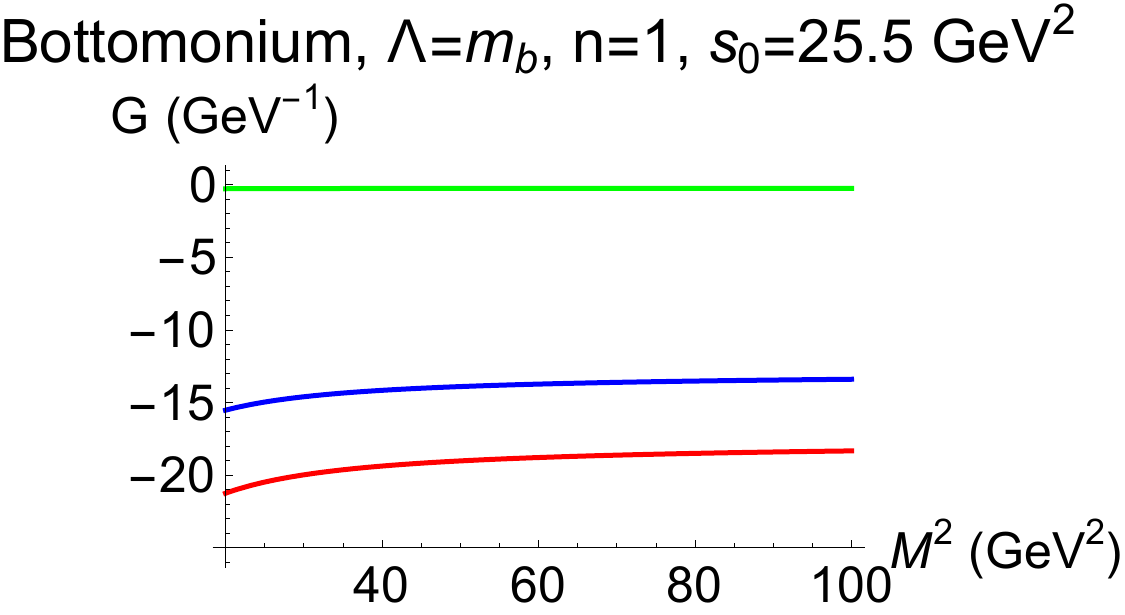}} \subfloat{\includegraphics[width=0.35\linewidth]{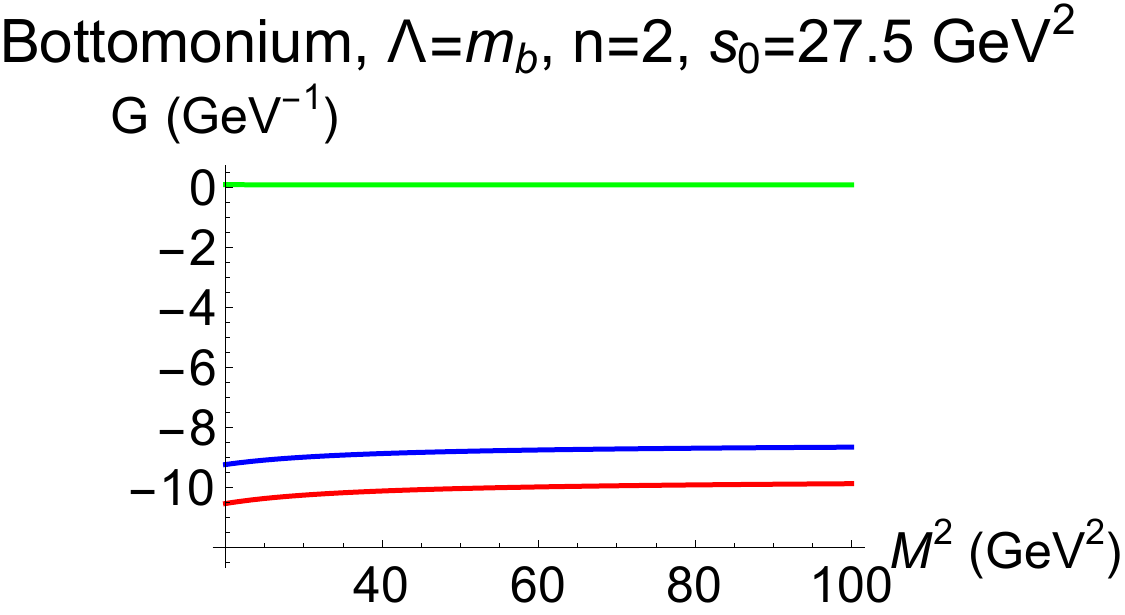}} \subfloat{\includegraphics[width=0.35\linewidth]{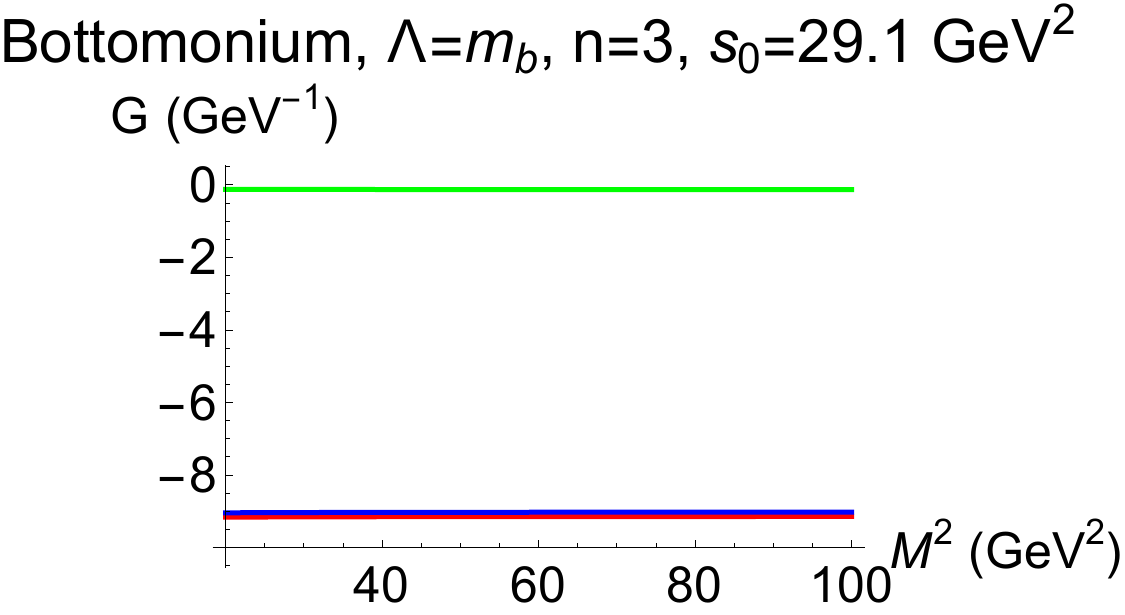}}
\\

    \caption{$G$ vs $M^2$. Red: $G_1 - G_2$. Blue: $G_1$. Green: $G_2$.}
    \label{GvsM}
\end{figure}

 





\begin{figure}[ht]

     \subfloat{\includegraphics[width=0.35\linewidth]{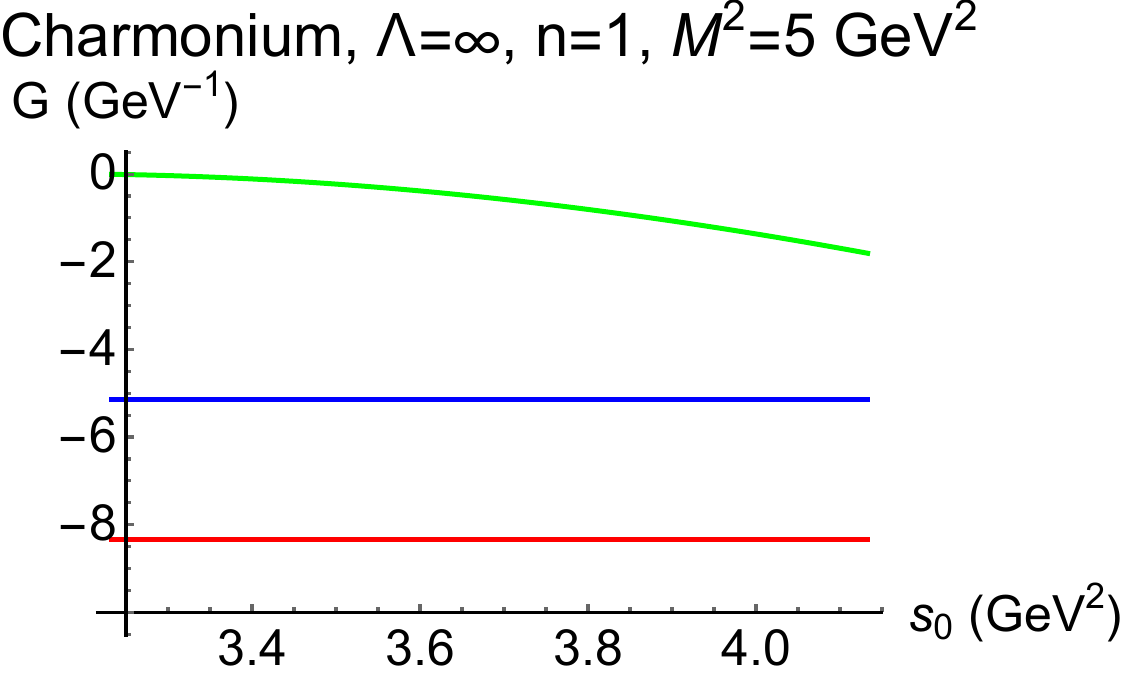}} \subfloat{\includegraphics[width=0.35\linewidth]{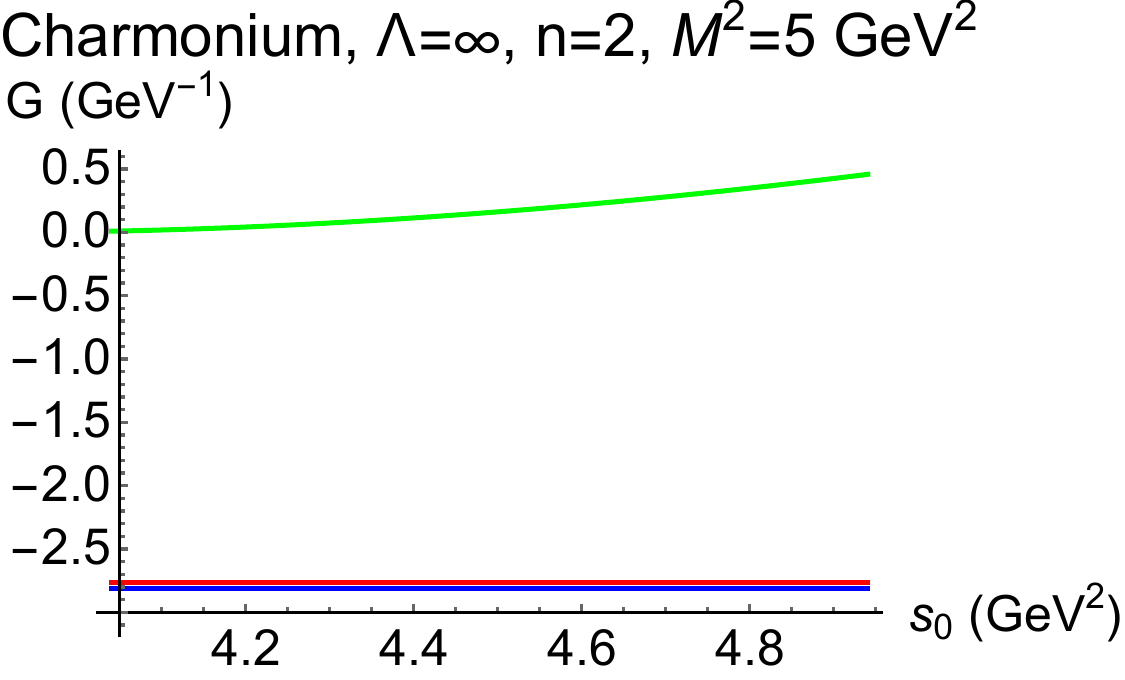}} \subfloat{\includegraphics[width=0.35\linewidth]{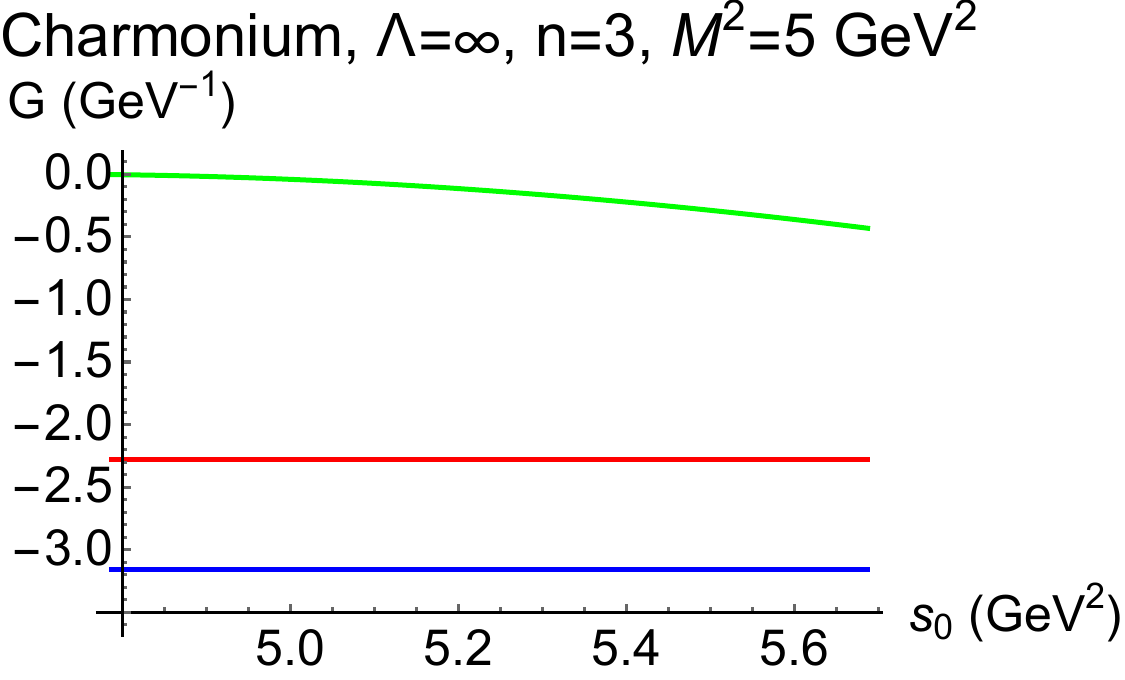}}
\\
     \subfloat{\includegraphics[width=0.35\linewidth]{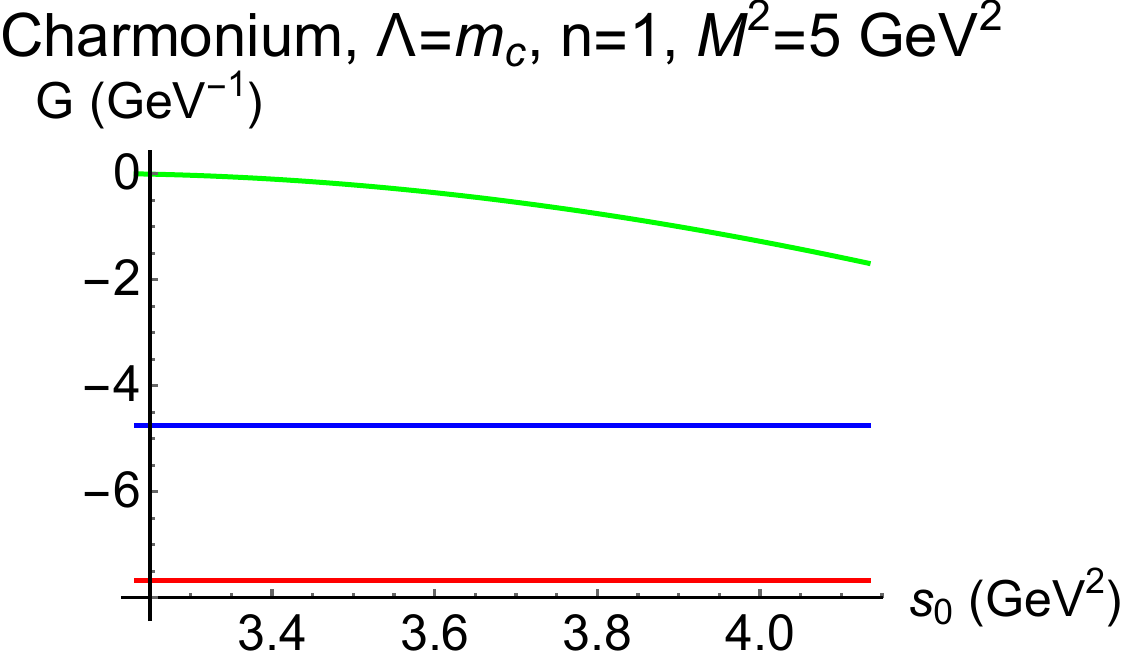}} \subfloat{\includegraphics[width=0.35\linewidth]{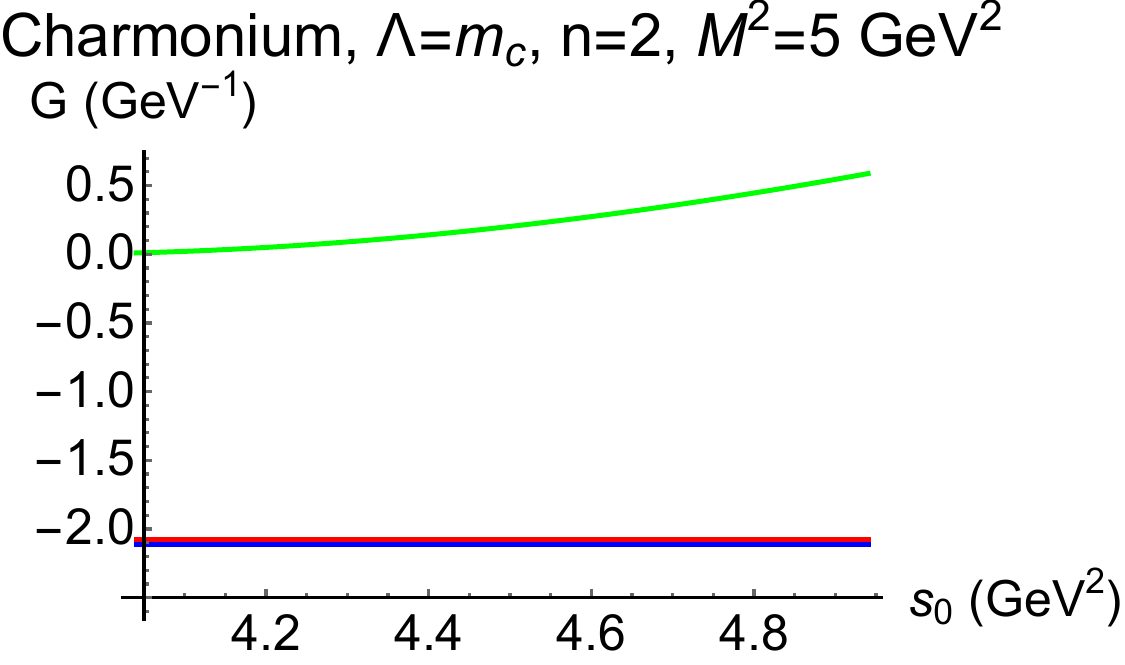}} \subfloat{\includegraphics[width=0.35\linewidth]{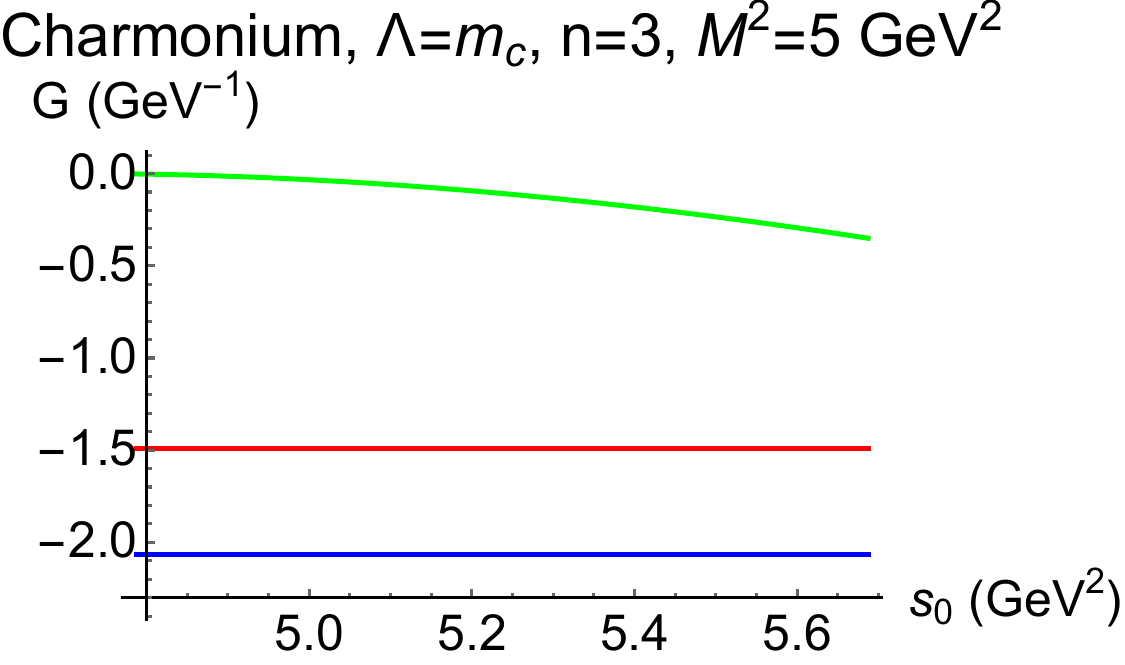}}
\\
     \subfloat{\includegraphics[width=0.35\linewidth]{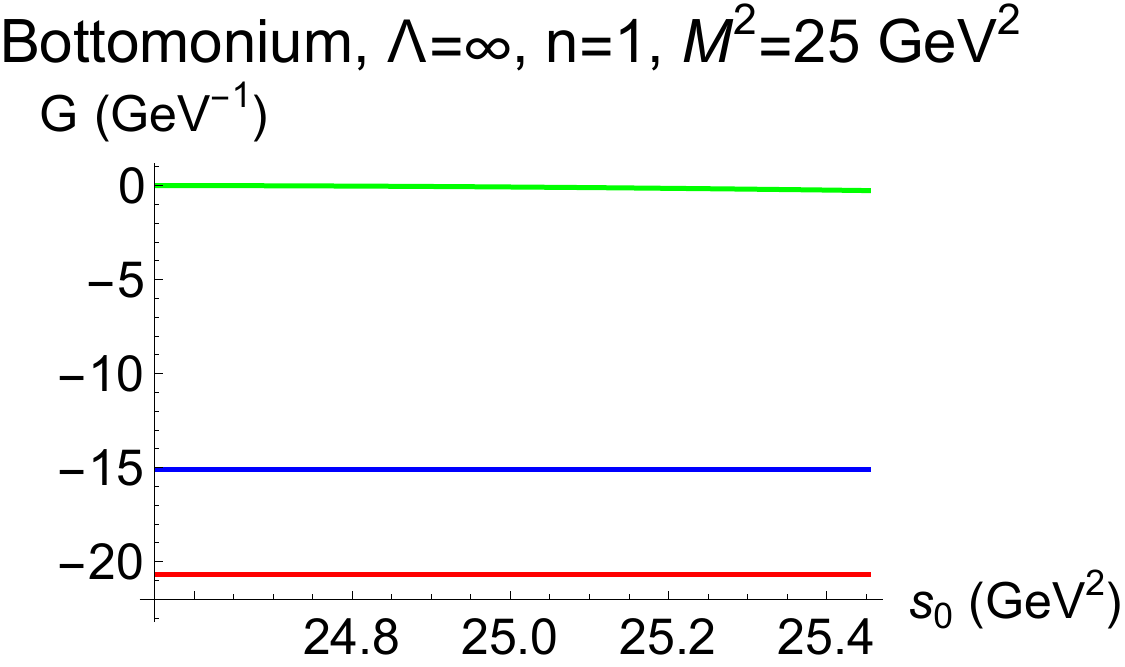}} \subfloat{\includegraphics[width=0.35\linewidth]{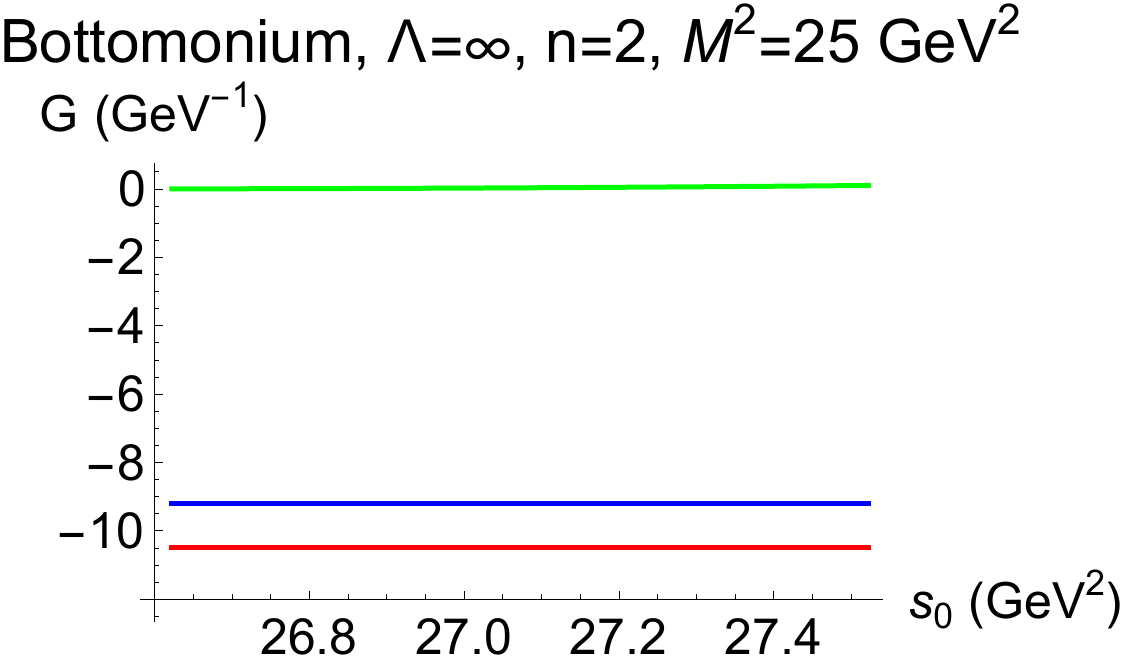}} \subfloat{\includegraphics[width=0.35\linewidth]{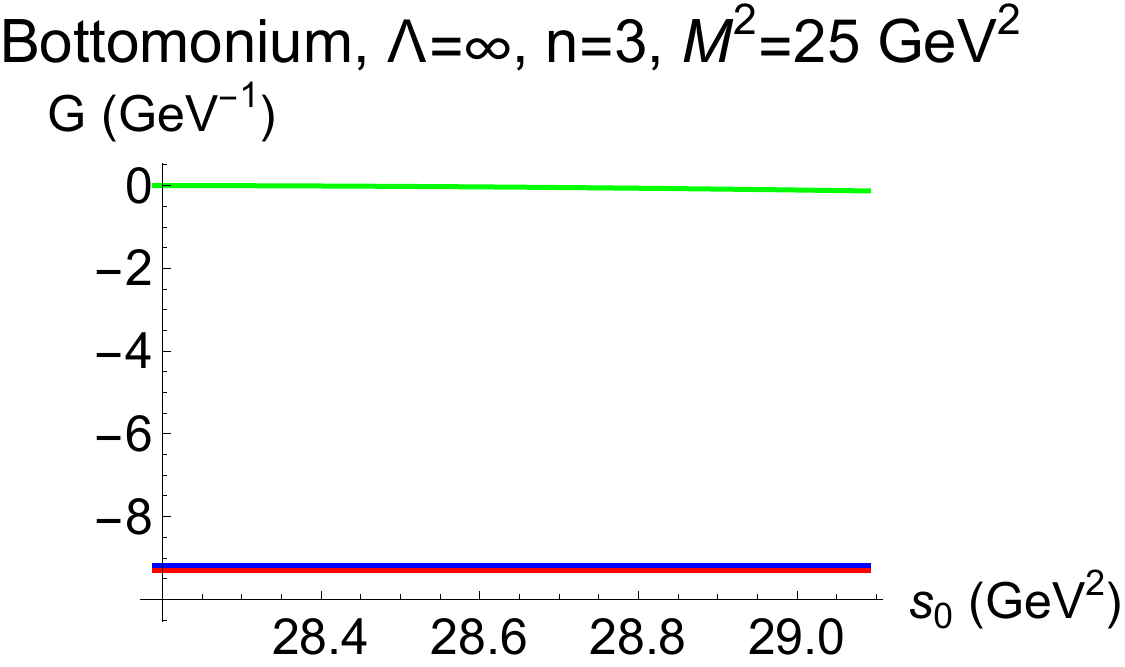}}
\\
     \subfloat{\includegraphics[width=0.35\linewidth]{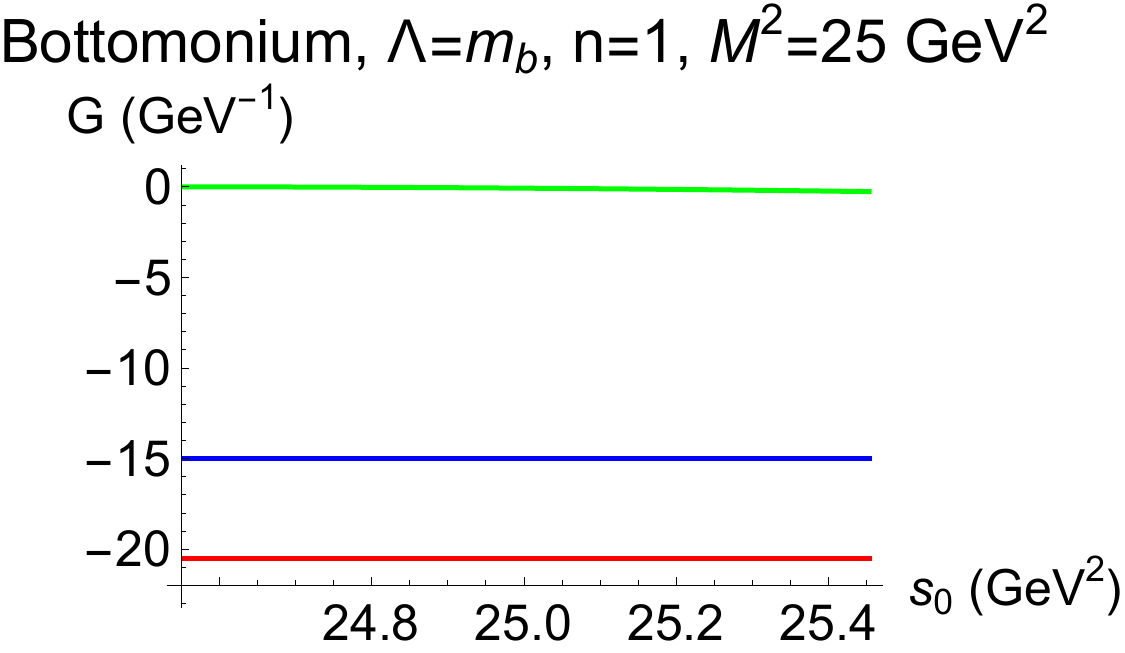}} \subfloat{\includegraphics[width=0.35\linewidth]{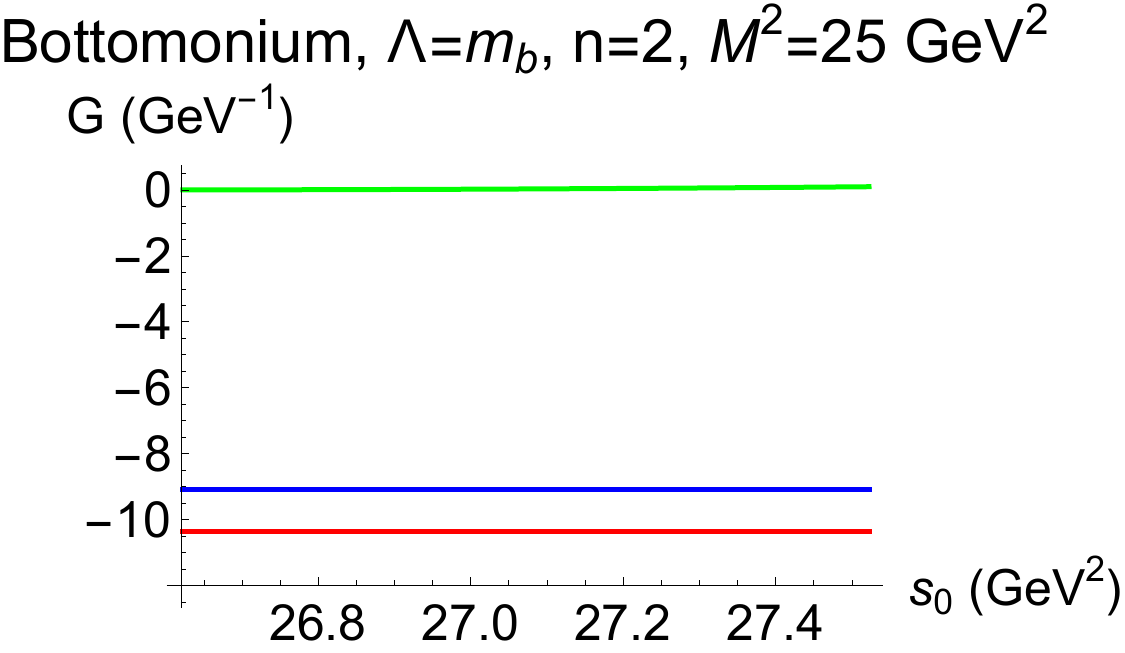}} \subfloat{\includegraphics[width=0.35\linewidth]{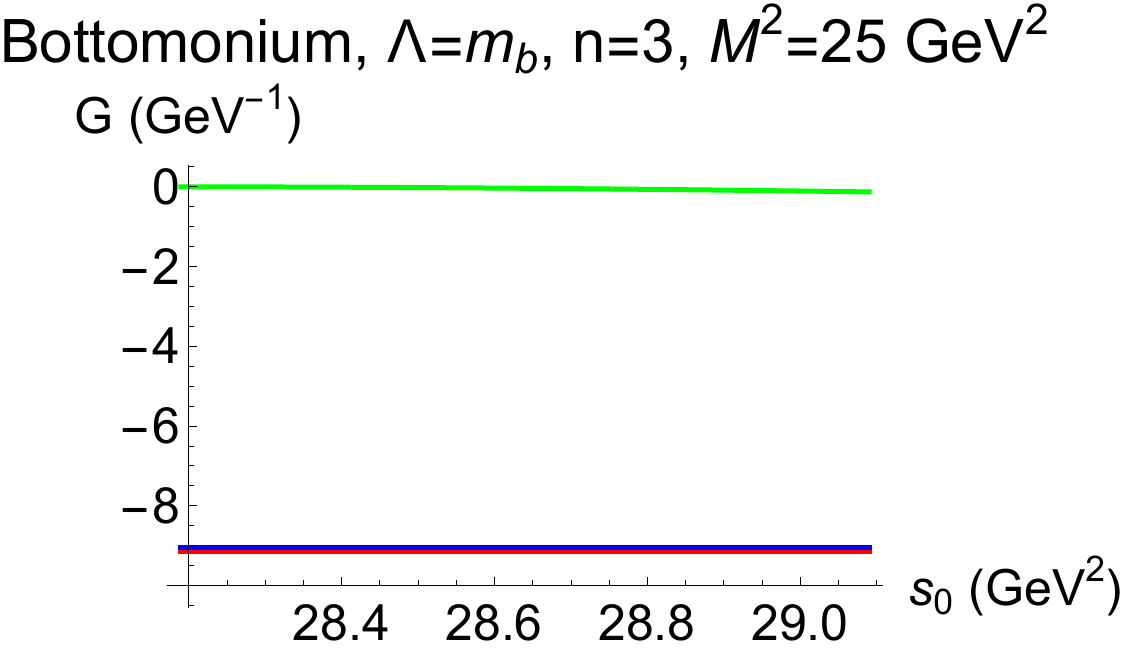}}
\\
    \caption{$G$ vs $s_0$. Red: $G_1 - G_2$. Blue: $G_1$. Green: $G_2$.}
    \label{GvsA}
\end{figure}


\begin{figure}[ht]
   
     \subfloat{\includegraphics[width=0.35\linewidth]{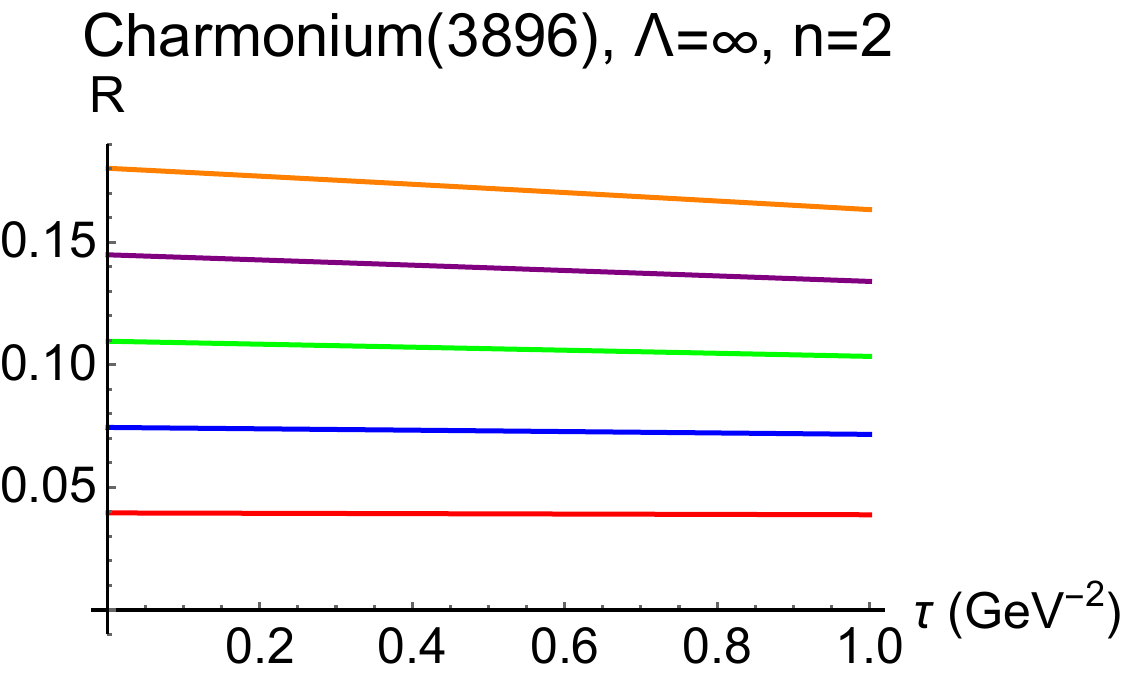}} \subfloat{\includegraphics[width=0.35\linewidth]{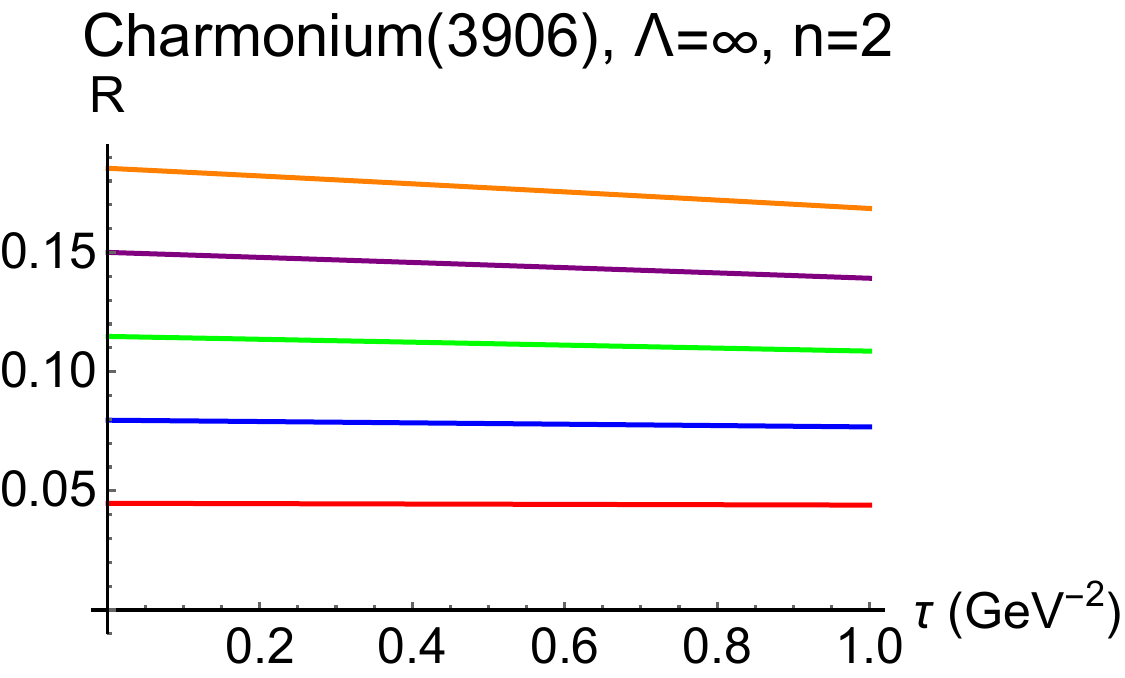}} \subfloat{\includegraphics[width=0.35\linewidth]{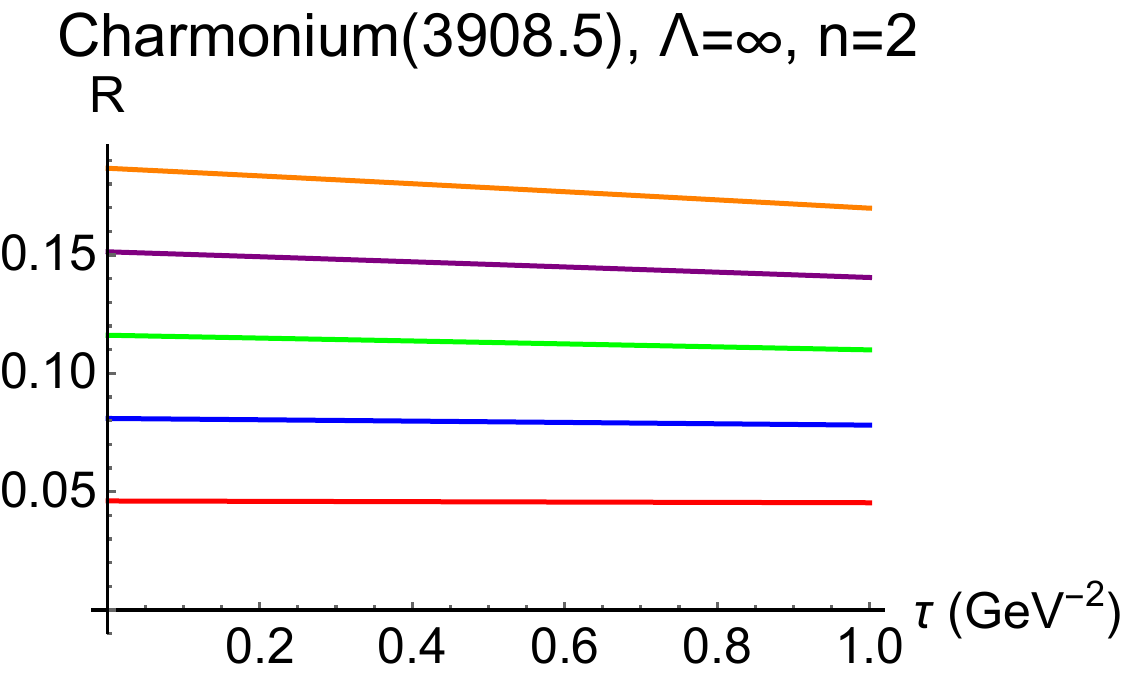}}
\\
     \subfloat{\includegraphics[width=0.35\linewidth]{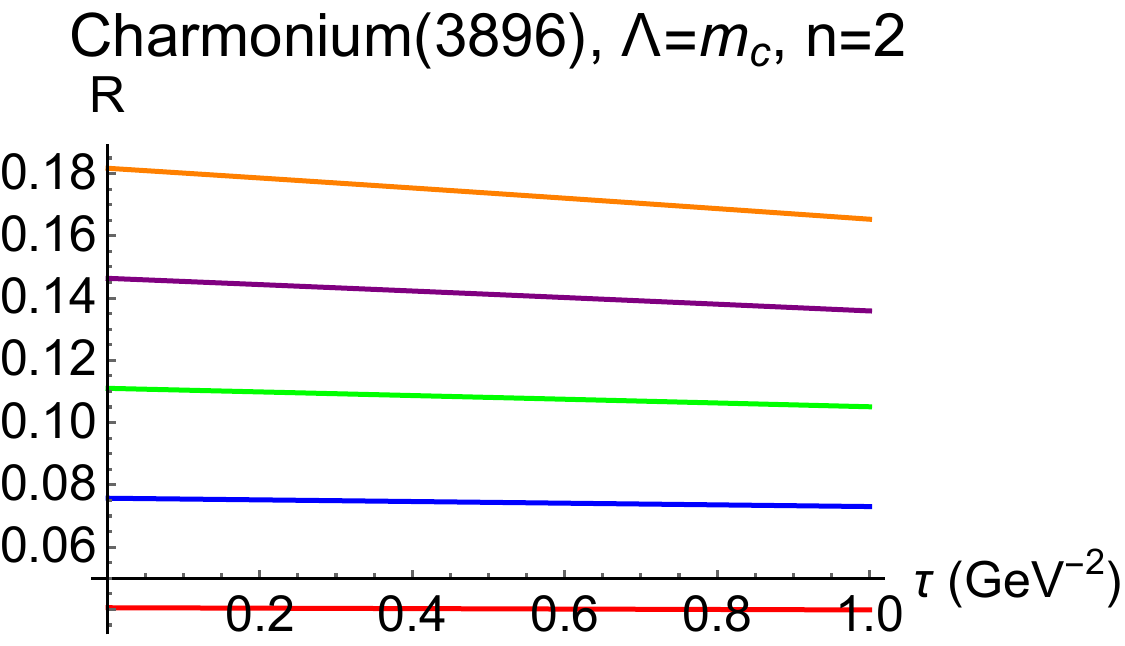}} \subfloat{\includegraphics[width=0.35\linewidth]{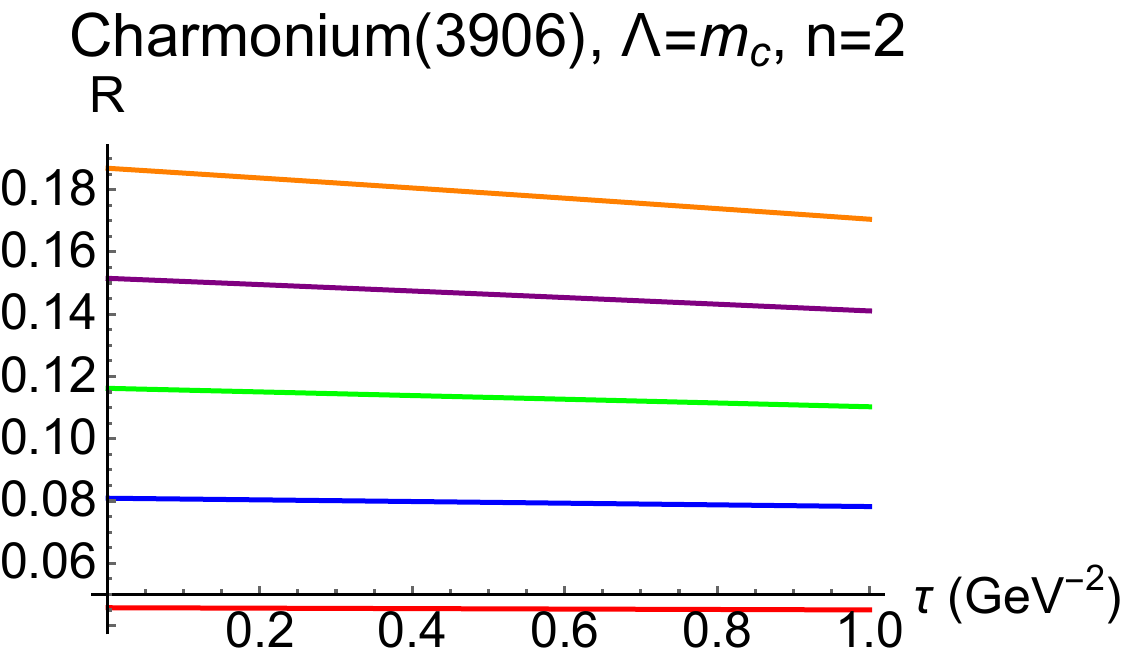}} \subfloat{\includegraphics[width=0.35\linewidth]{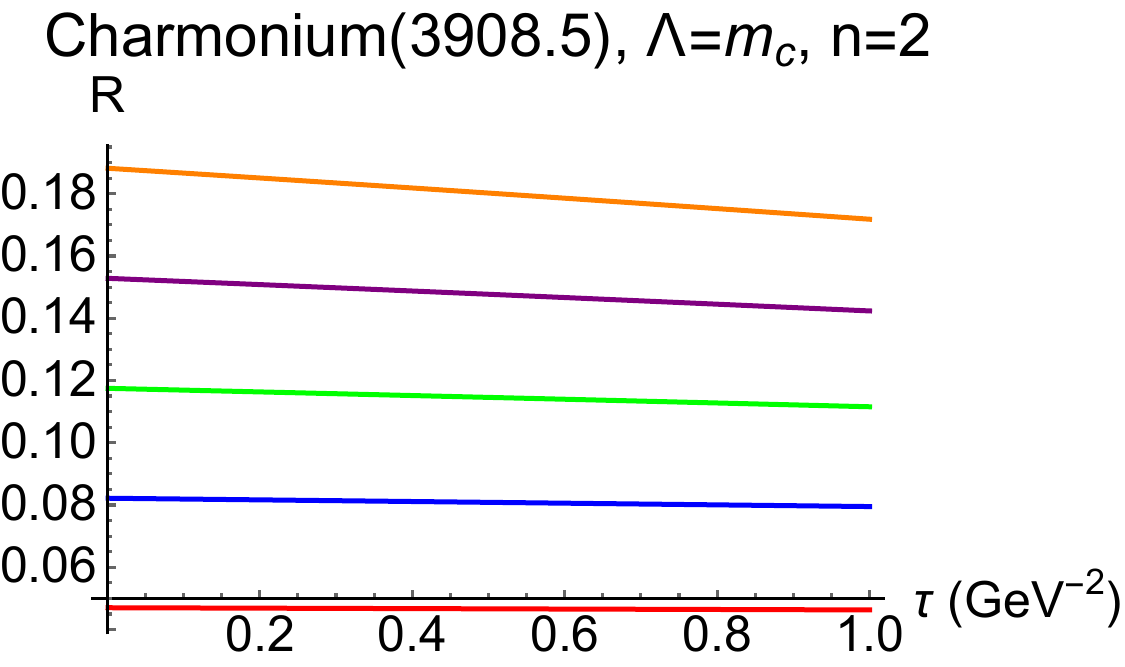}}
\\
    \caption{$R$ vs $\tau =1/M^2$ for $n=2$ charmonium states, where the charmonium mass is varied. Mass values are taken from \cite{Cincioglu2016} and indicated in parantheses on each plot. Again, $s_0 = \frac{m_{c\bar{c}}^2}{4}+\alpha $. Red: $\alpha = 0.2\, GeV^2$. Blue: $\alpha = 0.4\, GeV^2$. Green: $\alpha = 0.6\, GeV^2$. Purple: $\alpha = 0.8\, GeV^2$. Orange: $\alpha = 1.0\, GeV^2$. As seen in these plots, using the corresponding $m_{c\bar{c}}$ in the calculation of the couplings is still acceptable within our prescribed uncertainty range; that is, $R$ values are required to be less than 30\%.}
    \label{residualsnew}
\end{figure}